\documentclass[twoside,journal,10pt]{IEEEtran}
\usepackage{makecell}
\usepackage{hyperref}
\usepackage{array}
\usepackage{graphicx,amssymb,amsmath}
\usepackage{multicol}
\usepackage[noadjust]{cite}
\usepackage{setspace}
\usepackage{subfigure}
\usepackage{float}
\usepackage {url}
\usepackage{stfloats}
\usepackage{amsthm,pifont}
\usepackage{flushend}
\usepackage{cases,subeqnarray}
\usepackage{bm,multirow,bigstrut}
\usepackage{textcomp}
\usepackage{latexsym,bm}
\usepackage{booktabs}
\usepackage{xcolor}
\usepackage{mathtools}
\usepackage{dsfont}
\usepackage{extarrows}
\usepackage{epsfig}
\usepackage{epsfig}
\usepackage{epstopdf}
\usepackage{colortbl}
\usepackage{algpseudocode}
\usepackage{algorithmicx,algorithm}
\usepackage{fancyhdr}
\theoremstyle{plain}

\theoremstyle{plain}

\usepackage{amsmath}

\definecolor{Gray}{gray}{0.85}
\begin{document}
\title{DRESS: Diffusion Reasoning-based Reward Shaping Scheme For Intelligent Networks}
\author{Feiran You, Hongyang~Du$^*$, Xiangwang Hou, Yong~Ren,~\IEEEmembership{Senior Member,~IEEE}, and Kaibin~Huang,~\IEEEmembership{Fellow,~IEEE}
\thanks{F. You, H.~Du, and K.~Huang are with the Department of Electrical and Electronic Engineering, University of Hong Kong, Hong Kong SAR, China (email: fryou@eee.hku.hk, duhy@eee.hku.hk, huangkb@hku.hk).}
\thanks{X. Hou and Y. Ren are with the Department of Electronic Engineering, Tsinghua University. Y. Ren is also with the State Key laboratory of Space Network and Communications, Tsinghua University, Beijing 100084, China (e-mail: xiangwanghou@163.com, reny@tsinghua.edu.cn).}
}
\fancypagestyle{firstpage}{
\fancyhf{}  
\lhead{%
\begin{minipage}{0.9\textwidth}
\includegraphics[height=0.9cm]{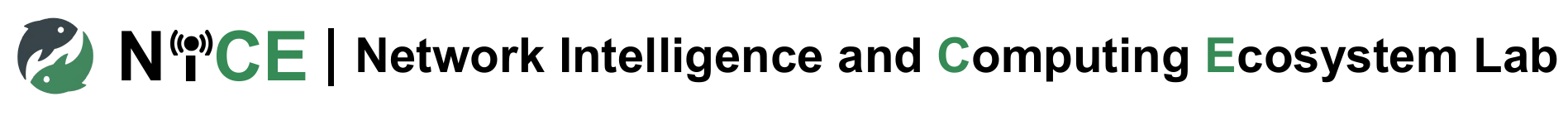}%
\end{minipage}
}
\renewcommand{\headrulewidth}{0pt}
}
\maketitle
\thispagestyle{firstpage}
\vspace{-1cm}
\begin{abstract}
Network optimization remains fundamental in wireless communications, with Artificial Intelligence (AI)-based solutions gaining widespread adoption. As Sixth-Generation (6G) communication networks pursue full-scenario coverage, optimization in complex extreme environments presents unprecedented challenges. The dynamic nature of these environments, combined with physical constraints, makes it difficult for AI solutions such as Deep Reinforcement Learning (DRL) to obtain effective reward feedback for the training process. However, many existing DRL-based network optimization studies overlook this challenge through idealized environment settings.
Inspired by the powerful capabilities of Generative AI (GenAI), especially diffusion models, in capturing complex latent distributions, we introduce a novel {\underline{D}}iffusion Reasoning-based {\underline{Re}}ward {\underline{S}}haping {\underline{S}}cheme (DRESS) to achieve robust network optimization.
By conditioning on observed environmental states and executed actions, DRESS leverages diffusion models' multi-step denoising process as a form of deep reasoning, progressively refining latent representations to generate meaningful auxiliary reward signals that capture patterns of network systems. Moreover, DRESS is designed for seamless integration with any DRL framework, allowing DRESS-aided DRL (DRESSed-DRL) to enable stable and efficient DRL training even under extreme network environments. 
Experimental results demonstrate that DRESSed-DRL achieves about $1.5{\rm{x}}$ times faster convergence than its original version in sparse-reward wireless environments and significant performance improvements in multiple general DRL benchmark environments compared to baseline methods. The code of DRESS is available at \url{https://github.com/NICE-HKU/DRESS}.
\end{abstract}

\begin{IEEEkeywords}
Diffusion model, reinforcement learning, wireless networks, reward reshaping, and optimization
\end{IEEEkeywords}
\IEEEpeerreviewmaketitle
\section{Introduction}
The deployment of Sixth-Generation (6G) wireless communications is pushing communication networks into extreme scenarios, e.g., underwater industrial monitoring, high-mobility unmanned aerial vehicle swarms, and disaster response scenarios, introducing significant demands on advanced network optimization~\cite{10061632,10364654,10239370}. 
However, unlike terrestrial networks, these environments exhibit unique challenges. For example, underwater acoustic channels suffer from severe multipath fading and propagation delays, while aerial networks face rapid topology changes and intermittent line-of-sight conditions. Modeling for these wireless environments typically relies on abstract mathematical fading models, e.g., Rayleigh fading, that inadequately capture the actual channel behaviors. Consequently, conventional optimization methods, which depend on these inaccurate models, are ill-equipped to handle the inherent system dynamics~\cite{10681174,10198239}.

\begin{figure}[t]
\centering
\includegraphics[width = 0.45\textwidth]{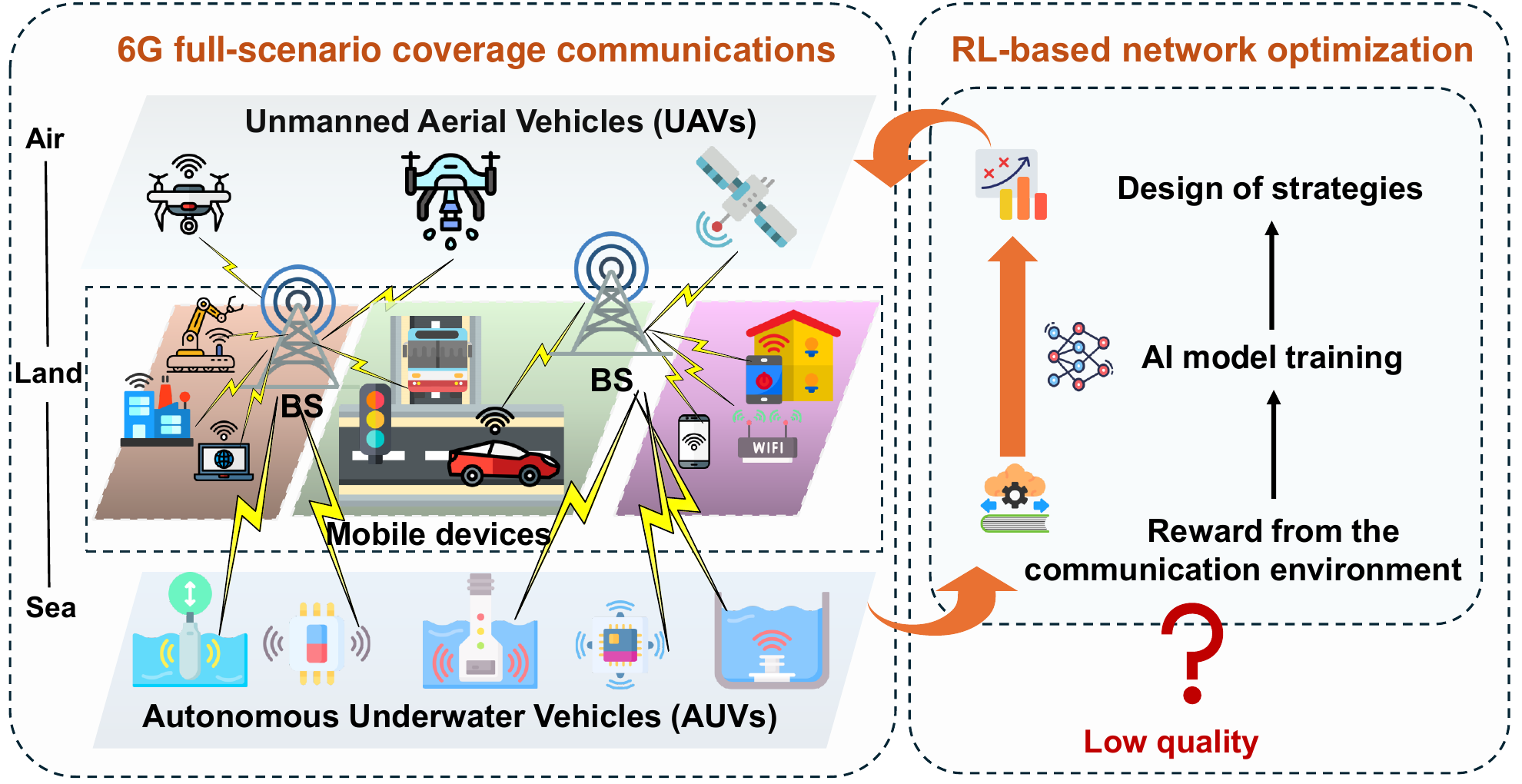}
\caption{Motivation: the limitation of the RL applied in the 6G full-scenario coverage communication system with low-quality reward feedback in varying environments.}
\label{fig_mo}
\end{figure}

Deep Reinforcement Learning (DRL) has gained attention as a candidate solution for extreme 6G scenarios due to its ability to self-adapt to environmental changes, handle high-dimensional state spaces, and optimize multiple performance objectives simultaneously~\cite{10552627}. However, the effectiveness of conventional DRL approaches is fundamentally limited in 6G complex communications system environments, where reward signals are often sparse, delayed, or even unavailable~\cite{cao2024survey}, as shown in Fig.~\ref{fig_mo}.
For instance, highly dynamic channel conditions and frequent link disruptions can result in infrequent or unreliable feedback, creating a significant mismatch with DRL's reliance on dense and instantaneous reward signals. Furthermore, the complex coupling between multiple system parameters exacerbates the challenge, as it expands the optimization space and makes optimal solutions increasingly rare~\cite{wang2022deep}. 
The sparse reward problem represents a critical bottleneck in applying DRL to extreme communication environments, where performance feedback is often delayed or incomplete, making it difficult to associate specific actions with system improvements. This scarcity of positive feedback, combined with the extensive parameter space in wireless network optimization, significantly reduces the learning efficiency of conventional DRL algorithms.

Several methods have been proposed to improve reward signals' quality in DRL, such as curiosity-driven exploration modules~\cite{wu2024securing}, reward shaping~\cite{eschmann2021reward}, and curriculum learning~\cite{xue2022using}. These approaches have also seen preliminary applications in communication networks.
For example, tailored reward functions in 6G network slicing~\cite{zhang2024digital} and hierarchical architectures in UAV swarm coordination~\cite{li2024reinforcement} have been proposed to guide DRL policies. 
While these methods help mitigate the challenge of training DRL in 6G communication networks, they face critical limitations.
Specifically, existing approaches often rely on task-specific reward engineering or require complex environmental interactions, which can be resource-intensive and difficult to scale. Additionally, their reward mechanisms are tightly coupled with specific network architectures, limiting their adaptability across diverse scenarios. For instance, a reward function optimized for aerial networks may not generalize well to underwater communication systems. This lack of cross-scenario flexibility highlights a fundamental drawback in current solutions. Furthermore, methods like reward shaping depend on prior knowledge or expert demonstrations, which are often limited, expensive to acquire, and may introduce human biases.
To overcome these limitations, we advocate for a universal solution with the following requirements:
\begin{itemize}
\item \textit{R1. Architecture-agnostic compatibility}: Seamless integration with existing DRL frameworks (e.g., DQN, SAC, GNN hybrids) without structural modifications
\item \textit{R2. Extreme-environment robustness}: Reliable pseudo-reward generation under sparse/delayed feedback through implicit modeling of state-action relationships
\end{itemize}

To address the two aforementioned requirements, we propose the {\underline{D}}iffusion Reasoning-based {\underline{Re}}ward {\underline{S}}haping {\underline{S}}cheme (DRESS). Inspired by the powerful modeling capabilities of diffusion models in the era of Generative AI (GenAI)~\cite{du2024enhancing}.
DRESS trains a diffusion model-based agent to automatically learn a dense auxiliary reward signal without requiring external human knowledge or introducing additional objectives. 
The multi-step denoising process inherent in diffusion models serves as a form of deep reasoning, progressively refining the latent representation of the current environmental state and the executed action to infer meaningful reward signals even when environmental feedback is limited. This generated auxiliary reward can then be combined with the environment's original reward to form the total reward signal for the training process of DRESS-aided DRL (DRESSed-DRL).

Unlike traditional reward shaping methods, DRESS introduces a separate diffusion model to generate reward signals without interfering with the DRL architecture and training process or imposing any additional architectural requirements on the DRL framework. Therefore, DRESS satisfies {\textit{R1 architecture-agnostic compatibility}} and can be used together with nearly all DRL algorithms.
Furthermore, compared to the very recent reward agent-based method for reward shaping~\cite{mareward}, the outperforming distribution modeling capabilities of the diffusion model-based reasoning processes are expected to generate higher-quality auxiliary reward signals. 
During the early stages of training, when environmental reward signals are sparse or of low quality, DRESS is expected to effectively assist the action network in exploration, thereby accelerating training convergence. 
In the later stages, when action quality improves, DRESS could help stabilize the training process and further enhance the final system performance. This approach addresses {\textit{R2 extreme environment robustness}} and provides a universal and efficient solution for applying DRL in complex communication networks.

The main contributions of this paper are as follows:

\begin{itemize}
\item We propose DRESS, a general-purpose architecture-agnostic reward shaping framework that integrates seamlessly with diverse DRL algorithms without requiring structural modifications, enabling deployment across heterogeneous network scenarios.
\item We leverage diffusion model-based reasoning as DRESS's core component to address reward sparsity challenges in dynamic wireless environments, autonomously generating high-fidelity auxiliary reward signals without human priors.
\item We validate DRESS through various experiments on a general wireless network optimization scenario and several DRL benchmark environments. The results show that DRESSed-DRL enhances exploration in the early training phases by inferring the quality of latent action from partial observations, achieving convergence about $1.5{\rm{x}}$ times faster compared to the baseline DRL. In later stages, DRESS stabilizes DRL training through adaptive reward calibration, showing $33\%$ performance gains compared to the advanced reward shaping method.
\end{itemize}
The remainder of this paper is organized as follows: Section~\ref{section2} discusses related work in DRL for wireless networks, reward shaping, and generative diffusion models. Section~\ref{environment} presents the wireless network optimization benchmark environment. Section~\ref{section4} details the design of DRESS and DRESSed-DRL algorithms. Section~\ref{secnums} presents comprehensive experimental results and analysis. Finally, Section~\ref{Cons} concludes the paper with a summary of our key findings.

\section{Related Work}\label{section2}
In this section, we discuss related works, including DRL for wireless networks, reward shaping, and generative diffusion models.

\subsection{DRL for Wireless Networks}
DRL has emerged as a powerful tool for optimizing wireless networks, offering data-driven solutions to complex challenges in resource allocation, network slicing, and dynamic control. Its ability to learn adaptive policies through environment interactions makes it particularly suited for stochastic wireless environments \cite{feriani2021single,nguyen2020deep}. Recent advances demonstrate DRL’s versatility across diverse applications, such as multi-agent coordination for decentralized resource allocation in UAV-assisted RANs \cite{tccn_ml1}, privacy-aware computation offloading in Mobile Edge Computing (MEC) networks using Multi-agent Deep Deterministic Policy Gradient (MADDPG) \cite{10763434}, and cross-layer optimization through the integration of diffusion models with DRL for joint ASP selection and resource allocation in AI-generated content (AIGC) networks \cite{10409284}. Additionally, survey works like \cite{10766420} systematically apply DRL to specialized environments such as underwater networks, addressing unique challenges in spectrum allocation and energy efficiency.
However, these methods fundamentally assume the availability of high-quality reward signals during policy learning—an assumption rarely valid in operational networks, especially in extreme conditions. Real-world wireless environments exhibit reward sparsity (e.g., delayed throughput feedback), reward ambiguity (e.g., correlated QoS metrics), and reward non-stationarity (e.g., user mobility patterns). This discrepancy creates a critical reality gap: policies trained on idealized reward models often degrade when deployed in practical systems. 
Recent efforts partially address these challenges through scenario-specific adaptations, such as employing diffusion models to enhance ASP selection rewards in AIGC networks \cite{10409284} or designing custom reward functions for MEC privacy constraints \cite{10763434}. However, these solutions remain tightly coupled with specific DRL architectures or application domains, failing to address the compounded challenges of delayed feedback, partial observability, and non-stationary interference. 
This gap motivates a general method that enhances the robustness of DRL optimization in complex and extreme wireless environments.

\subsection{Reward Shaping}
Reward shaping provides a systematic approach to enhance DRL training by supplementing sparse environmental feedback with auxiliary signals, yet its application in wireless networks remains underexplored. 
While foundational studies like \cite{grzes2017reward} proved its ability to accelerate DRL convergence, recent approaches follow divergent paths: imitation-based methods infer rewards from expert trajectories \cite{perrusquia2024uncovering}, rule-driven strategies embed domain knowledge via policy entropy \cite{10196410} or safety constraints \cite{10571558}, and hybrid architectures such as \cite{9797851} integrate shaped rewards with actor-critic frameworks for energy management.
Despite these advancements, several main challenges remain in wireless network optimization. 
First, existing methods often depend on task-specific DRL architectural modifications (e.g., customized critic networks) or human-crafted heuristics, limiting their adaptability to various DRL algorithms and heterogeneous network topologies. 
Second, while intrinsic reward agents reduce the need for manual tuning, they struggle to model complex state-action relationships in highly dynamic and interference-heavy wireless environments. This often leads to sensitivity to hyperparameters and poor generalization, especially when feedback is delayed or observations are incomplete. These issues highlight the need for a universal reward shaping framework that can automatically generate accurate rewards without requiring specific architectures or prior knowledge.


\subsection{Diffusion Models}
Diffusion models, exemplified by pioneering architectures like GLIDE~\cite{nichol2022glide}, DALL·E-2~\cite{dalle2}, and Stable Diffusion~\cite{stabdiff}, have revolutionized Generative Artificial Intelligence (GenAI) by synthesizing photorealistic outputs through iterative denoising. Unlike traditional methods such as Generative Adversarial Networks (GANs) and Variational Autoencoders (VAEs), which often struggle with mode collapse or blurred outputs, diffusion frameworks—including Denoising Diffusion Probabilistic Models (DDPM)~\cite{ho2020denoising} and Denoising Diffusion Implicit Models (DDIM)~\cite{song2020denoising}—progressively refine noisy inputs into high-fidelity results by reversing a predefined noise injection process. This unique mechanism enables precise control over generation dynamics, making them particularly adept at capturing complex data distributions.
This capability has driven their adoption in wireless networks for two key roles: (1) {\textit{network design}}, where they act as semantic decoders in 6G semantic communication systems to reconstruct bandwidth-efficient multimedia data, e.g., medical imaging under low signal-to-noise ratios~\cite{du2023generative}, and (2) {\textit{network optimization}}, where diffusion-enhanced DRL algorithms model dynamic environments for tasks like multi-cell interference management.
However, despite their success in these roles, diffusion models remain unexplored for a critical task: leveraging their latent distribution modeling power to generate auxiliary rewards that universally enhance the convergence speed and robustness of any DRL architecture—especially in extreme wireless scenarios with delayed feedback, partial observability, or unpredictable interference.


\section{Wireless Benchmark Environment}\label{environment}
\begin{figure}[t]
\centering
\includegraphics[width=0.45\textwidth]{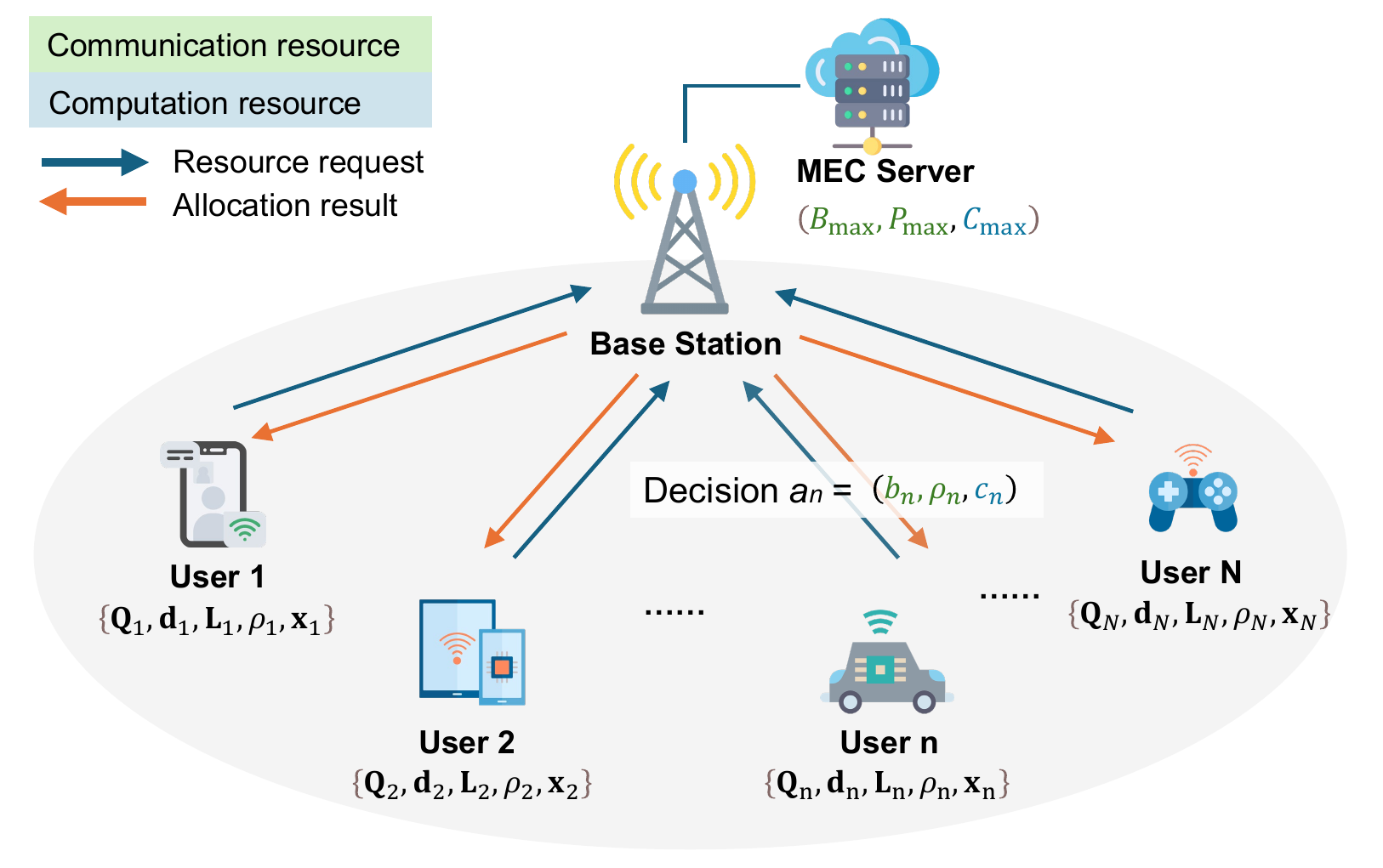}
\caption{{\textit{MECLatency}}: one centralized BS in management of the communication and computation resources and $N$ users proposing resource requests.}
\label{fig_sys}
\end{figure}

In this section, we design a multi-resource allocation problem in wireless networks as the benchmark environment to evaluate DRL algorithms and the proposed DRESS framework. 
To demonstrate extreme wireless conditions, we use the complexity of MEC problems and their stringent latency requirements to simulate scenarios with missing reward signals.


\subsection{Wireless Environment}
We consider a MEC scenario where a Base Station (BS) serves $N$ mobile users, as shown in Fig.~\ref{fig_sys}.
Specifically, the BS manages three types of resources: bandwidth $B \in [0, B_{\text{max}}]$, transmit power $P \in [0, P_{\text{max}}]$, and computational capacity $C \in [0, C_{\text{max}}]$. Each user $n \in \{1,\dots,N\}$ requests service with specific requirements defined by computational demand $d_n$ (in CPU cycles), maximum tolerable latency $L_n$, and minimum throughput threshold $T_n$.
The position of the $n_{\rm{th}}$ user evolves according to a bounded random walk model with velocity $v_n$ and direction $\theta_n$. At each time step $t$, the position update as
\begin{equation}
\mathbf{x}_n(t+1) = \min(\max(\mathbf{x}_n(t) + \Delta \mathbf{x}_n(t), 0), 1),
\end{equation}
where $\Delta{\mathbf{x}_n(t)}$ represents random displacement drawn from a uniform distribution $\mathcal{U}(-\Delta x_{\text{max}}, \Delta x_{\text{max}})$. The min-max operations constrain user positions within the normalized BS coverage radius, where 0 represents the BS location and 1 represents the cell edge. Each user has a patience threshold $\rho_n$, and if service requirements are not met within this threshold, the user leaves the system, resulting in a significant decrease in the Quality of Service (QoS).

The time-varying dynamic channel quality for each user is updated based on the new position and environmental factors, creating a challenging environment for resource allocation optimization. We consider the wireless channel to incorporate both large-scale path loss and small-scale Rayleigh fading. For the $n_{\rm{th}}$ BS-user link with distance $D_n$, the received signal power can be expressed as
\begin{align}
{{\bf{r}}_n}  = & \:\: {p_n}{P_{{\rm{max}}}}\sqrt {D_n^{ - {\alpha _n}}} {{\bf{H}}_n}{{\bf{w}}_n}{s_n} + {{\bf{n}}_n}
\notag\\& + \sum\limits_{j \ne n} {{p_j}} {P_{{\rm{max}}}}\sqrt {D_j^{ - {\alpha _j}}} {{\bf{H}}_j}{{\bf{w}}_j}{s_j},
\end{align}
where $p_n \in [0, 1]$ is the power allocation fraction for the $n_{\rm{th}}$ user, $P_{{\rm{max}}}$ is the total transmit power, $\mathbf{H}_n \in \mathbb{C}^{1 \times M}$ represents the Rayleigh fading channel matrix, $\mathbf{w}_n = \mathbf{H}_n^{\rm H}/||\mathbf{H}_n||_F$ denotes the MRC weight vector, $s_n$ is the unit-energy transmitted symbol, and $\mathbf{n}_n \sim \mathcal{CN}(0,\sigma^2\mathbf{I})$ models additive white Gaussian noise. The path loss component $D_n^{-\alpha}$ depends on transmission distance $D_n$ and the path loss exponent $\alpha$.

In multi-user scenarios, the Signal-to-Interference-plus-Noise Ratio (SINR) for the $n_{\rm{th}}$ user can be expressed as
\begin{equation}
\text{SINR}_n = \frac{{p_n}{P_{{\rm{max}}}} D_n^{-\alpha} \sum_{m=1}^M |h_{n,m}|^2}{ \sum_{j\neq n} {p_j}{P_{{\rm{max}}}} D_j^{-\alpha} \sum_{m=1}^M |h_{j,m}|^2 + \sigma^2},
\end{equation}
where $h_{n,m}$ denotes the complex channel coefficient for the $j_{\rm{th}}$ path. Under Rayleigh fading assumptions, $|h_{n,m}|^2$ follows an exponential distribution, making the combined channel gain $\sum_{j=1}^M |h_{n,m}|^2$ gamma-distributed with shape parameter $M$.

Subsequently, in this wireless benchmark environment, several key performance indicators can be derived. For the $n_{\rm{th}}$ user, the achievable throughput follows Shannon's capacity formula as
\begin{equation}
T_n = b_n B_{\text{max}} \log_2(1 + \text{SINR}_n),
\end{equation}
where $b_n \in [0, 1]$ is the bandwidth allocation fraction for the $n_{\rm{th}}$ user. Considering the communications and computing delay, the service latency is
\begin{equation}
L_n = \frac{d_n}{T_n} + \frac{d_n}{c_n C_{\text{max}}},
\end{equation}
where $c_n \in [0, 1]$ is the computational resource allocation fraction for the $n_{\rm{th}}$ user. The corresponding service rate is determined by
\begin{equation}
S_n = \frac{1}{L_n},
\end{equation}
where $d_n$ represents the data demand in bits.

\subsection{Service-rate Oriented Optimization Problem}
We consider a service-rate oriented optimization problem, aiming to maximize $\sum_{n=1}^{N} S_n$.
However, the wireless network system operates under three key requirements that create optimization challenges:
\begin{itemize}
\item {\textit{Strict Latency Requirements:}} Each user $n$ demands minimum throughput $T_n^{\text{req}}$ and maximum tolerable latency $L_n$ for services like video streaming and edge computing. The BS must maintain these metrics while serving multiple users with limited resources.
\item {\textit{Dynamic Channel Conditions:}} The channel quality varies with user mobility and environmental factors, making resource allocation decisions time-sensitive. Poor allocation decisions can lead to service interruptions and user departures when the patience threshold $\rho_n$ is exceeded.
\item {\textit{Resource Constraints:}} The BS must optimize the allocation of limited bandwidth $B_{\text{max}}$, transmit power $P_{\text{max}}$, and computational capacity $C_{\text{max}}$ across all users while maintaining system stability.
\end{itemize}

From the AI-based solution perspective, $\sum_{n=1}^{N} S_n$ serves as the primary reward signal for optimization. However, in 6G full-scenario coverage, obtaining reliable reward signals from the environment, i.e., $r^{\left( E\right)}$, becomes particularly challenging due to the aforementioned challenges. For example, service performance measurement may become impossible in latency-critical scenarios like 6G-aided industrial control networks. Where robotic arms perform precision assembly, once latency surpasses 1 ms, emergency stops are triggered to prevent equipment damage, making performance measurement infeasible. 
The challenge is further amplified in dynamic environments where network conditions change rapidly. In underwater communication networks, signal propagation paths are constantly altered by ocean currents and marine activities. When acoustic channels experience severe multi-path fading, receivers cannot decode signals correctly, leading to complete communication blackouts.

To simulate the phenomenon of reward signal loss in our wireless environment benchmark, we use latency as the trigger condition and thus name our environment {\textit{MECLatency}}. Specifically, the environment reward signal $r_E$ has two cases:

\begin{itemize}
\item \textbf{Case 1 (No Feedback):} When the current latency exceeds the critical latency threshold, i.e., $L_{n} > L_{th}$ $\left( {\exists n \in N} \right)$, we have 
\begin{equation}\label{rewardeq1}
r^{\left( E\right)} = 0.
\end{equation}
The system enters an outage state where no reliable performance metrics can be obtained. This represents practical scenarios such as connection timeouts or task failures, where the actual system performance becomes unmeasurable.

\item \textbf{Case 2 (Degraded Feedback):} When the latency is within the latency threshold, i.e., $L_{n} \leq L_{\rm{th}}$, the system receives delayed and incomplete feedback. The reward is:
\begin{equation}\label{rewardeq3}
r^{\left( E\right)} = \sum_{n=1}^{N} S_n - \lambda \sum_{n=1}^{N} \left( {{L_n} - {L_{\rm{th}}}} \right) - \sum_{n=1}^{N} \mu_n,
\end{equation}
where $\lambda$ and $\mu_n$ are the penalty factors, and $\mu_n$ can be expressed as
\begin{equation}
{\mu _n} = \left\{ {\begin{array}{*{20}{l}}
\mu, \qquad {\rm{if}} \:\: {\rho_n} \le {\rho_{{\rm{th}}}}\\
0.
\end{array}} \right.
\end{equation}
\end{itemize}

By considering the resource allocation for $N$ users, where the decision variables include:
bandwidth allocation vector ${\bf{b}} = \left\{ {{b_1},\ldots,{b_N}} \right\}$, transmit power allocation vector ${\bf{p}} = \left\{ {{p_1},\ldots,{p_N}} \right\}$, and computational resource allocation vector ${\bf{c}} = \left\{ {{c_1},\ldots,{c_N}} \right\}$, the optimization goal in {\textit{MECLatency}} is to maximize the reward function as 
\begin{equation}\label{rewardsignal}
\begin{aligned}
\max_{\left\{ {\textbf{b}}, {\textbf{p}}, {\textbf{c}} \right\}} \quad & r_E =
\begin{cases}
\eqref{rewardeq1}, \qquad \text{(Case 1)} \\
\eqref{rewardeq3}, \qquad \text{(Case 2)} \\
\end{cases} \\
\text{s.t.,} \quad 
& \sum_{n=1}^{N} b_{n} \leq 1, \sum_{n=1}^{N} p_{n} \leq 1, \sum_{n=1}^{N} c_{n} \leq 1, \\
& b_{n}, p_{n}, c_{n} \in [0, 1], \: \forall n \in N.
\end{aligned}
\end{equation}
For this wireless benchmark environment, we can increase the optimization difficulty by tightening the latency requirement, i.e., reducing $L_{\rm th}$. When $L_{\rm th}$ is set to a small value, many DRL-based resource allocation algorithms keep receiving zero rewards that provide no meaningful learning signals for the DRL agent training, making convergence difficult or impossible. We further examine this challenge and our proposed solution in Section~\ref{secnums}.

\section{Diffusion Reasoning-based Reward Shaping Scheme}\label{section4}
\begin{figure}[t]
\centering
\includegraphics[width=0.48\textwidth]{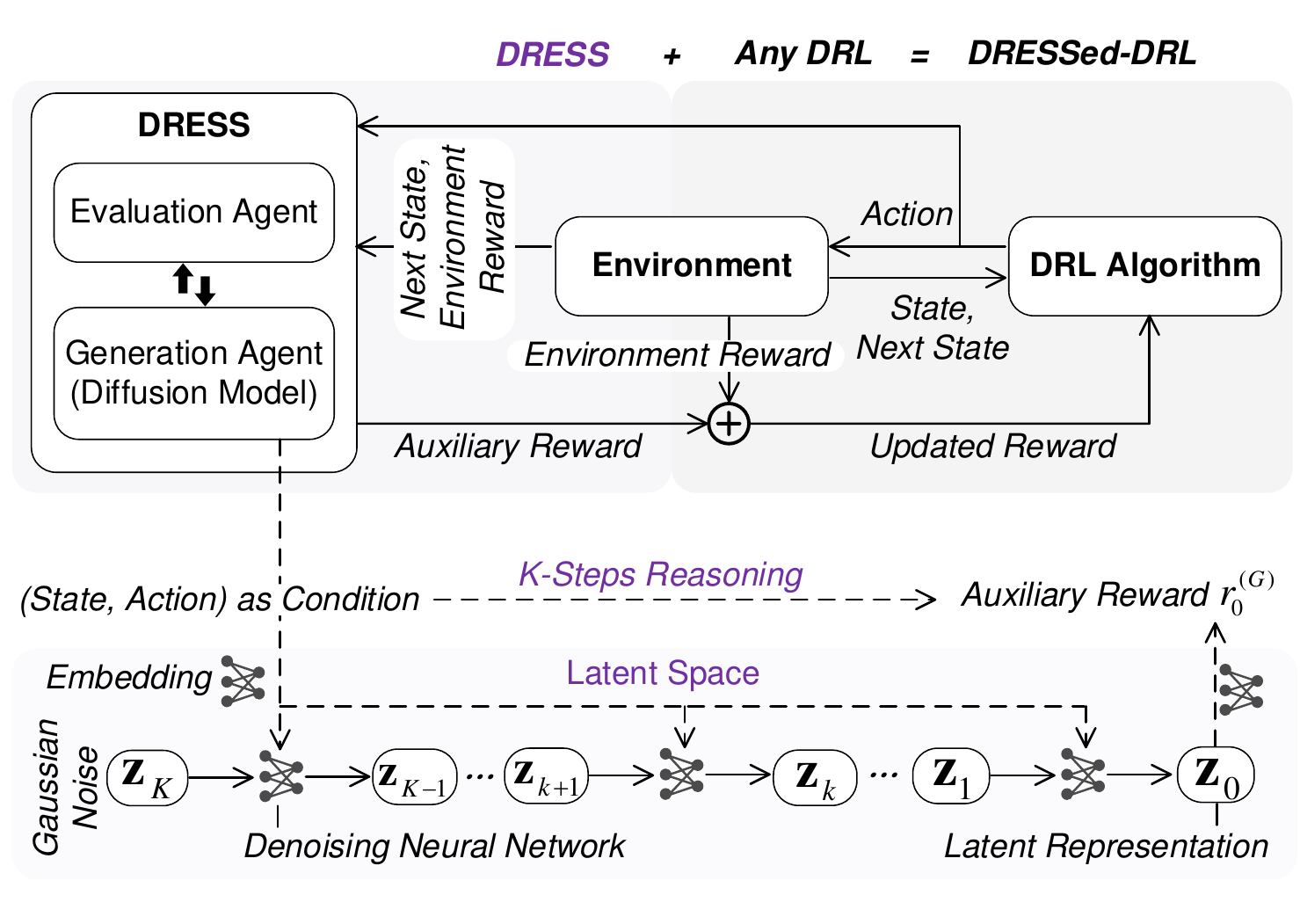}
\caption{The proposed DRESS framework and its integration with various DRL architectures.}
\label{fig_algo}
\end{figure}

In this section, we introduce DRESS, beginning with a formal definition of a DRL framework for our considered wireless benchmark environment, followed by the DRESS architecture and training methodology. Finally, we discuss the universal integration of DRESS with arbitrary DRL algorithms.

\subsection{Formal DRL Framework Definition}

DRL is a learning framework that enables sequential decision-making through direct interactions with an environment $\mathcal{E}$. At each timestep $t$, a state $s_t \in \mathcal{S}$ is observed from the state space, based on which an action $a_t \in \mathcal{A}$ is selected according to the learned strategy. The environment responds with a sparse reward signal $r_t^E \in \mathbb{R}$ and transitions to the next state $s_{t+1}$ through the transition dynamics $T: \mathcal{S} \times \mathcal{A} \rightarrow \mathcal{S}$. Therefore, the wireless network optimization problem is formulated as a Markov Decision Process (MDP) $(\mathcal{S}, \mathcal{A}, \mathcal{T}, \mathcal{R}^E, \gamma)$ with:

{\textit{Environment dynamics $\mathcal{E}$}} includes dynamics like user mobility, throughput calculation, and patience decrement. It also contains constraints on bandwidth, power, and computational capacity. The agent observes the following parameters from the environment at each time step $t$: the time-varying channel conditions and user mobility; resource constraints including bandwidth $B_{total}$, power $P_{\max}$, and compute $C_{\max}$; and user behavior models incorporating throughput demands and patience decay.

\textit{State Space $\mathcal{S}$} captures instantaneous network status at each time step $t$, denoted as $\mathbf{s}_t$, as
\begin{equation}
{\bm{s}_t} = \left\{ {{{\bf{Q}}_t},{{\bf{d}}_t},{{\bf{L}}_t},{\rho _t},{{\bf{x}}_t}} \right\},
\end{equation}
where ${{\bf{Q}}_t}$  is the channel quality states that can be expressed as
\begin{equation}
{{\bf{Q}}_t} = \left\{ {\sqrt {D_{1,t}^{ - {\alpha _{1,t}}}} {{\bf{H}}_{1,t}}{{\bf{w}}_{1,t}}, \ldots,\sqrt {D_{N,t}^{ - {\alpha _{N,t}}}} {{\bf{H}}_{N,t}}{{\bf{w}}_{N,t}}} \right\},
\end{equation}
${\bf{d}}_t = \left\{ d_1,\ldots,d_N\right\}$ represents the user demands, latency requirements are given by ${{\bf{L}}_t} = \left\{ L_1, \ldots,L_N \right\}$, patience levels are denoted as ${\bm{\rho}_t} = \left\{ \rho_1, \ldots,\rho_N \right\}$, and spatial positions are expressed as ${{\bf{x}}_t} = \left\{ x_1, \ldots,x_N \right\}$.

\textit{Action space $\mathcal{A}$} defines resource allocation decisions for $N$ users:
\begin{align}
{{\bm{a}}_t} & = \left\{ {{{\bm{a}}_1},\ldots,{{\bm{a}}_N}} \right\}
\notag\\&
= \left\{ {\left\{ {{b_1},{p_1},{c_1}} \right\},\ldots,\left\{ {{b_N},{p_N},{c_N}} \right\}} \right\}.
\end{align}
where $b_i \in [0,1]$, $p_i \in [0,1]$, and $c_i \in [0,1]$ represent bandwidth, power, and compute allocations respectively.

\textit{Environment reward $r_t^{(E)}$} represents the feedback signal from the environment, i.e., \eqref{rewardsignal}, after executing action ${{\bm{a}}_t}$ in state ${{\bm{s}}_t}$ at time step $t$. For most DRL approaches, the objective is to maximize the expected cumulative discounted reward:
\begin{equation}
J = \mathbb{E}\left[\sum_{t=0}^{\infty} \gamma^t r_t^{(E)}\right],
\end{equation}
where $\gamma \in [0,1)$ is the discount factor. This objective manifests differently across DRL paradigms: in value-based methods like DQN, it is achieved by learning an optimal action-value function; in policy-based methods, a parameterized policy $\pi_\theta$ directly maximizes $J$; while actor-critic methods combine both approaches by simultaneously learning a policy and a value function.

However, traditional DRL methods often struggle with sparse reward signals, particularly in complex network environments where the immediate environmental reward $r_t^{(E)}$ frequently returns zero values, providing limited information for learning. In our wireless network benchmark environment, we simulate this challenging scenario by imposing strict latency requirements in a dynamic setting. The sparsity of reward signals creates a credit assignment problem - meaning the agent has difficulty identifying which specific actions in a sequence contributed to eventual successes or failures - making it challenging for traditional DRL methods to connect individual actions with their long-term impacts.

\subsection{DRESS based Auxiliary Reward Generation}
To address the inherent limitations of sparse environmental reward signals in complex network optimization, we propose DRESS, which generates complementary auxiliary signals. Traditional reward shaping methods struggle in dynamic wireless environments where critical state transitions often occur without immediate reward feedback. DRESS overcomes this challenge by learning latent reward patterns through the joint analysis of network states and allocation actions. The core innovation lies in conditioning the diffusion process on state-action pairs $\left( {\bm{s}}_t,{\bm{a}}_t\right) $ to generate auxiliary rewards $r_t^{(G)}$ that capture system dynamics not encoded in $r_t^{(E)}$. By jointly considering network states and actions, DRESS learns to evaluate the potential long-term value of allocation decisions, distinguishing it from conventional diffusion models that require clean target samples.

\subsubsection{Diffusion Model Training Process}
The diffusion process consists of two phases, the forward diffusion phase and the reverse denoising phase~\cite {du2024enhancing}.
First, we consider an idealized accurate assessment of state-action pairs $\left( {\bm{s}}_t,{\bm{a}}_t\right) $ as auxiliary reward signals, denoted as $r_t^{(G)}$, which can enable DRL models to train effectively in environments with extremely sparse rewards.
In the forward diffusion process, auxiliary reward signals are progressively corrupted by Gaussian noise over a series of $K$ time steps. Let $r_{k,t}^{(G)}$ denote the data at diffusion time-step $k$ and DRL time-step $t$, where $r_{0,t}^{(G)} \equiv r_{t}^{(G)}$ represents the actual data used at DRL time-step $t$. At each diffusion time-step, a Gaussian noise with a variance of $\beta_k$ is added to ${\bf{x}}_{k-1}$ to yield ${\bf{x}}_k$ with the distribution $q\left(r_{k,t}^{(G)}|r_{k-1,t}^{(G)}\right)$. This process can be expressed mathematically as:
\begin{equation}
q\left(\! {\left. {r_{k,t}^{(G)}} \right|{r_{k-1,t}^{(G)}}} \right)\! = {\cal N}\left( {r_{k,t}^{(G)}; {{\bm{\mu}}_k},{{\bf{\Sigma }}_k} } \right),
\end{equation}
where $q\left( {\left. {{r_{k,t}^{(G)}}} \right|{r_{k-1,t}^{(G)}}} \right)$ is a normal distribution, characterized by the mean ${{\bm{\mu }}_k}  = \sqrt {1 - {\beta _t}} {r_{k-1,t}^{(G)}}$ and the variance ${\bf{\Sigma}}_k = {\beta _k}{\bf{I}}$, and ${\bf{I}}$ is the identity matrix indicating that each dimension has the same standard deviation ${\beta _k}$.

When $K$ is large, ${r_{K,t}^{(G)}}$ approximates an isotropic Gaussian distribution~\cite{ho2020denoising}. By learning the reverse distribution $q\left({\left.{r_{k-1,t}^{(G)}}\right|{r_{k,t}^{(G)}}}\right)$, we can sample ${r_{K,t}^{(G)}}$ from $\mathcal{N}\left(\mathbf{0},\mathbf{I}\right)$, execute the reverse process with $K$ denoising steps, and obtain $q\left({r_{0,t}^{(G)}}\right)$ as the auxiliary reward signal for the $t_{\rm th}$ DRL time-step. Since directly computing $ q\left( {\left. {{r_{k-1,t}^{(G)}}} \right|{{r_{k,t}^{(G)}}}} \right) $ is intractable, we approximate it with a parameterized model ${p_\theta}$ as follows:
\begin{equation}
p_\theta \!  \left(\! {r_{k-1,t}^{(G)}}|{r_{k,t}^{(G)}}\right) \!\!  =\!  \mathcal{N} \!  \left(\!  {r_{k-1,t}^{(G)}}; {\bm \mu}_\theta({r_{k,t}^{(G)}},k),{\bm \Sigma}_\theta({r_{k,t}^{(G)}},k)\right).
\end{equation}
Subsequently, we obtain the trajectory from ${r_{K,t}^{(G)}}$ to ${r_{0,t}^{(G)}}$ as
\begin{equation}
{p_\theta }\left( {{r_{0:K,t}^{(G)}}} \right) = {p_\theta }\left( {{{r_{K,t}^{(G)}}}} \right)\prod\limits_{t = 1}^T {{p_\theta }} \left( {{r_{k-1,t}^{(G)}} \mid {r_{k,t}^{(G)}}} \right).
\end{equation}
We define $ {\alpha _t} = 1 - {\beta _t} $ and $ {{\bar \alpha }_t} = \prod\limits_{j = 0}^t {{\alpha_j}} $. After adding the condition information, e.g., state-action pairs $\left( {\bm{s}}_t,{\bm{a}}_t\right) $, in the denoising process, ${p_\theta({r_{k-1,t}^{(G)}}|{r_{k,t}^{(G)}},{\bm{s}}_t,{\bm{a}}_t)}$ can be modeled as a noise prediction model with the covariance matrix fixed as~\cite{ho2020denoising}
\begin{equation}
{\bf{\Sigma}}_\theta\left({r_{k,t}^{(G)}}, {\bm{s}}_t,{\bm{a}}_t, t\right)=\beta_t \bf{I},
\end{equation}
and the mean is constructed as
\begin{equation}
{{\bm{\mu }}_\theta } \!  \left(\!  {{{r_{k,t}^{(G)}}},{\bm{s}}_t,{\bm{a}}_t,k} \! \right) = \frac{ {{r_{k,t}^{(G)}} - \frac{{{\beta_k}}}{{\sqrt {1 - {{\bar \alpha }_k}} }}{{\bm{\epsilon}} _\theta }\left(\!  {{r_{k,t}^{(G)}},{\bm{s}}_t,{\bm{a}}_t,k} \right)}}{{\sqrt {{\alpha_k}} }}.
\end{equation}
We first sample $ {r_{K,t}^{(G)}} \sim\mathcal{N}({\bm{0}},{\bm{I}}) $ and then form the reverse diffusion chain parameterized by $ \theta  $ as
\begin{equation}\label{denoise}
{{r_{k-1,t}^{(G)}}}\mid {{r_{k,t}^{(G)}}} = \frac{{{r_{k,t}^{(G)}}}}{{\sqrt {{\alpha _k}} }} - \frac{{{\beta _k}{{\bm{\epsilon}} _\theta }\left( {{{r_{k,t}^{(G)}}}, {\bm{s}}_t,{\bm{a}}_t,k} \right)}}{{\sqrt {{\alpha _k}\left( {1 - {{\bar \alpha }_k}} \right)} }} + \sqrt {{\beta _k}} {\bm{\epsilon}},
\end{equation}
where ${\bm{ \epsilon}} \sim\mathcal{N}({\bm{0}},{\bm{I}}) $ and $k = 1,\ldots,K$. Then, the loss function for the training process can be expressed as~\cite{ho2020denoising}
\begin{equation}\label{faef}
{{\cal L}} = {{\mathbb{E}}_{{r_{0,t}^{(G)}},k,{\bm{ \epsilon}}}}\left[ {{{\left\| {{\bm{ \epsilon}} - {{\bm{ \epsilon}}_\theta }\left( {\sqrt {{{\bar a}_k}} {r_{0,t}^{(G)}} + \sqrt {1 - {{\bar a}_k}} {\bm{ \epsilon}},{\bm{s}}_t,{\bm{a}}_t,k} \right)} \right\|}^2}} \right].
\end{equation}
However, the training process described above relies on the availability of idealized auxiliary reward signals $r_t^{(G)}$. In practical settings, we cannot directly access the optimal $r_t^{(G)}$ for arbitrary state-action pairs. A viable approach is to leverage the diffusion model's implicit reasoning capabilities to actively learn and infer the most appropriate auxiliary reward for each state-action pair.

\subsubsection{Diffusion Reasoning}
In each step of DRESS, we conduct reasoning within a latent space for the generation process. 
Unlike conventional approaches that directly predict a scalar reward, our method generates a latent representation which is then transformed into a auxiliary reward signal using maximum entropy principles. Additionally, for the optimal latent space representation ${\bm{z}_{0,t}^{(G)}}$ corresponding to the auxiliary reward signal ${r_{0,t}^{(G)}}$, we directly predict ${\bm{z}_{0,t}^{(G)}}$ rather than predicting the noise in each denoising step as shown in~\eqref{denoise}.
These design choices mitigate the instability often associated with direct prediction approaches, providing more consistent training dynamics~\cite{song2020denoising}. They also align with our diffusion reasoning paradigm, where the quality of the predicted reward ${r_{0,t}^{(G)}}$ will be evaluated and used as guidance for the diffusion model training~\cite{chi2023diffusion}.

To formalize our approach, we start with the posterior distribution for the reverse diffusion process:
\begin{equation}
q(\bm{z}_{k-1,t} | \bm{z}_{k,t}, \bm{z}_{0,t}) = \mathcal{N}(\bm{z}_{k-1,t}; \tilde{\mu}_k(\bm{z}_{k,t}, \bm{z}_{0,t}), \tilde{\beta}_k \mathbf{I}).
\end{equation}

The mean of this posterior can be expressed as:
\begin{equation}
\tilde{\mu}_k(\bm{z}_{k,t}, \bm{z}_{0,t}) = \frac{\sqrt{\alpha_{k-1}}(1-\bar{\alpha}_k)}{1-\bar{\alpha}_{k-1}}\bm{z}_{0,t} + \frac{\sqrt{\alpha_k}\beta_{k-1}}{1-\bar{\alpha}_{k-1}}\bm{z}_{k,t}.
\end{equation}

By replacing the true $\bm{z}_{0,t}$ with model's prediction $\hat{\bm{z}}_{0,\theta}(\bm{z}_{k,t}, \bm{s}_t, \bm{a}_t, k)$, we obtain the sampling equation as
\begin{align}\label{denoise2}
\bm{z}_{k-1,t}|\bm{z}_{k,t} = \: & \frac{\sqrt{\alpha_{k-1}}(1-\bar{\alpha}_k)}{1-\bar{\alpha}_{k-1}}\hat{\bm{z}}_{0,\theta}(\bm{z}_{k,t}, \bm{s}_t, \bm{a}_t, k) \notag \\
& + \frac{\sqrt{\alpha_k}\beta_{k-1}}{1-\bar{\alpha}_{k-1}}\bm{z}_{k,t} + \sigma_k\bm{\epsilon}.
\end{align}

Each step in this reverse diffusion process can be interpreted as a form of reasoning about the optimal latent representation for auxiliary rewards. As we iterate from $k=K$ down to $k=1$, the model progressively refines its understanding of the appropriate latent structure, effectively performing multi-step reasoning about the value of different state-action pairs.

To transform the final latent representation $\bm{z}_{0,t}$ into an actual reward signal, we employ a maximum entropy approach. Following the principles from soft actor-critic algorithms, we model the auxiliary reward distribution to balance expected reward and exploration:
\begin{align}
\bm{\kappa}_\phi &= f_{\mu}(\bm{z}_{0,t}), \\
\log \bm{\sigma}_\phi &= \text{clip}(f_{\sigma}(\bm{z}_{0,t}), \sigma_{\text{min}}, \sigma_{\text{max}}),
\end{align}
where $f_{\mu}$ and $f_{\sigma}$ are neural networks that map the latent representation $\bm{z}_{0,t}$ to the parameters of a normal distribution, with $\bm{\kappa}_\phi$ representing the mean of the auxiliary reward distribution and $\bm{\sigma}_\phi$ controlling its variance. The clipping operation, i.e.,  $ \text{clip}\left( \cdot\right)$, limits the log standard deviation between $\sigma_{\text{min}}$ and $\sigma_{\text{max}}$ to ensure numerical stability and prevent entropy collapse. The auxiliary reward is then sampled using the reparameterization trick as
\begin{equation}\label{denoise3}
r_{t}^{(G)} = \tanh(\bm{\kappa}_\phi + \bm{\sigma}_\phi \odot \bm{\epsilon}') \cdot e_s + e_b,
\end{equation}
where $r_{t}^{(G)} \equiv r_{0,t}^{(G)}$ represents the generated auxiliary reward signal, $\bm{\epsilon}' \sim \mathcal{N}(\bm{0}, \mathbf{I})$ is a standard normal noise vector that enables stochastic sampling, $e_s$ and $e_b$ are scaling and bias terms that map the normalized outputs to the appropriate reward range, and $\odot$ denotes element-wise multiplication.  The hyperbolic tangent function, i.e., $\tanh\left( \cdot\right)$, constrains the outputs to the range $[-1, 1]$ before scaling, ensuring bounded rewards and improving training stability

This entropy-regularized approach offers several advantages. it maintains exploration in the reward space while providing meaningful signals, ensures the generated auxiliary rewards remain informative yet stable through the bounded nature of the tanh transformation, and helps prevent reward exploitation while promoting robust learning.

To establish a direction for the diffusion reasoning process, we consider the goal-oriented optimization objective. Specifically, an \textit{evaluation network} $Q_\upsilon$ assigns a Q-value to each combination of state-action pair and generated auxiliary reward signal. This $Q_\upsilon$ network acts as a guidance tool for the training of the DRESS network. The objective for the auxiliary reward signal generation is to obtain latent representations that, when transformed into rewards, maximize the expected Q-value as
\begin{equation}\label{actortrain}
\mathop{\arg\min}\limits_{\theta, \phi} \mathcal{L}(\theta, \phi) = -\mathbb{E}_{r_{t}^{(G)}\sim \eqref{denoise2},  \eqref{denoise3}}\left[Q_\upsilon\left(\bm{s}_t,\bm{a}_t,r_{t}^{(G)}\right)\right].
\end{equation}

The evaluation network $Q_\upsilon$ learns to estimate the long-term value of a state-action-auxiliary reward combination $({\bm{s}}_t,{\bm{a}}_t,r_{0,t}^{(G)})$ through Temporal Difference (TD) learning. Intuitively, this network answers the question: {\textit{``How valuable is this auxiliary reward signal for achieving high cumulative returns in the future?''}} At each step, we compute a TD target as
\begin{equation}
y_t = r_t^{(E)} + \gamma Q_{\upsilon'}\left({\bm{s}}_{t+1},{\bm{a}}_{t+1},{{r_{0,t+1}^{(G)}}}\right),
\end{equation}
where $r_t^{(E)}$ represents the immediate environmental reward at the $t_{\rm th}$ DRL time-step, $\gamma \in [0,1)$ is the discount factor balancing immediate and future rewards, and the second term estimates future returns using a target generation evaluation network $Q_{\upsilon'}$ whose parameters are updated through soft updates. The Q-network is then optimized by minimizing the mean squared error between its predictions and these TD targets:
\begin{equation}\label{qualitytrain}
\mathop{\arg\min}\limits_{\upsilon} \mathcal{L}_Q(\upsilon) = {\mathbb{E}} \left[ {{{\left(y_t - {Q_{\upsilon}}\left({\bm{s}}_t,{\bm{a}}_t,{{r_{0,t}^{(G)}}}\right)\right)}^2}} \right].
\end{equation}
This training process enables the Q-network to learn accurate value estimates that can then guide the DRESS network in generating meaningful auxiliary rewards.

We observe that DRESS's training process can be formulated as a Partially Observable Markov Decision Process (POMDP), as the reward generation network only perceives state-action pairs $({\bm{s}}_t,{\bm{a}}_t)$ through the DRL agent's environmental interactions, without accessing the complete environment state.  
In this POMDP formulation, the global state space $\mathcal{S}^{G}$ encompasses the environmental state space $\mathcal{S}^{\mathcal{E}}$, action space $\mathcal{A}^{\mathcal{E}}$, and the policy's parameter space $\Theta$ (representing the DRL agent's internal parameters, e.g., Q-network weights or actor-critic parameters). Moreover, DRESS's observation space remains constrained to $\Omega^{G} = \mathcal{S}^{\mathcal{E}} \times \mathcal{A}^{\mathcal{E}}$, independent of the DRL agent's specific architecture.

This partial observability creates a universal adaptation mechanism where DRESS indirectly guides any DRL agent by generating auxiliary rewards $r^{(G)}$ that modify the agent's learning signals. The environmental rewards $r^{(E)}$ and generated rewards $r^{(G)}$ are combined as:
\begin{equation}\label{combinere}
r_t^{\text{total}} = r_t^{(E)} + \beta \cdot r_t^{(G)},
\end{equation}  
where $\beta$ is a scaling factor that controls the weight of the auxiliary reward signal. 

\subsubsection{DRESS Implementation}
Based on the loss functions in DRESS, i.e., \eqref{actortrain} and \eqref{qualitytrain}, we can identify two distinct agents: the {\textit{generation agent}} parameterized by $\theta$ and $\phi$ that are jointly optimized through backpropagation, and the {\textit{evaluation agent}} parameterized by $\upsilon$. We discuss how to integrate DRESS with DRL algorithms and address key design considerations.

{\textbf{Integration with DRL Algorithms:}}
This formulation ensures DRESS's architecture-agnostic nature, making it compatible with various DRL paradigms:
\begin{itemize}
\item For \textit{value-based methods} (e.g., DQN), $\Theta$ corresponds to Q-network parameters updated through temporal difference learning;
\item For \textit{policy-based methods} (e.g., PPO), $\Theta$ represents actor-network parameters optimized via policy gradients;
\item For \textit{actor-critic hybrids}, $\Theta$ encompasses both components.
\end{itemize}
For instance, in the Soft Actor-Critic (SAC) algorithm, which is an actor-critic method, the critic loss for the soft Q-function is computed as:
\begin{equation}\label{dressedsac}
\mathcal{L}_Q(\phi) = \mathbb{E}_{({\bm{s}}_t, {\bm{a}}_t, {\bm{s}}_{t+1}, r_t^{\text{total}}) \sim \mathcal{D}_P} \left[ \frac{\left(Q_\upsilon({\bm{s}}_t, {\bm{a}}_t) - Q_T\right)^2}{2} \right],
\end{equation}
where $Q_T = y(r_t^{\text{total}}, {\bm{s}}_{t+1})$ is the target value that can be calculated using the target Q-network $Q_{\upsilon'}$ as:
\begin{align}
Q_T = &  \gamma \mathbb{E}_{{\bm{a}}_{t+1} \sim \pi_\theta} \left[Q_{\upsilon'}({\bm{s}}_{t+1}, {\bm{a}}_{t+1}) - \alpha \log \pi_\theta({\bm{a}}_{t+1}|{\bm{s}}_{t+1})\right] \notag \\
& + r_t^{\text{total}}.
\end{align}
The actor aims to minimize the following loss:
\begin{equation}
\mathcal{L}_{\pi}(\theta) = \mathbb{E}_{{\bm{s}}_t \sim \mathcal{D}_P} \left[ \mathbb{E}_{{\bm{a}}_t \sim \pi_\theta} \left[ \alpha \log \pi_\theta({\bm{a}}_t|{\bm{s}}_t) - Q_\upsilon({\bm{s}}_t, {\bm{a}}_t) \right] \right].
\end{equation}
This example demonstrates how DRESS seamlessly integrates with existing DRL frameworks by augmenting their reward signals while maintaining their original optimization objectives. The integration is further enhanced through flexible implementation choices in both architecture and the training process.

{\textbf{Network Architecture Design:}}
The architecture of the denoising network can be flexibly chosen based on the structural characteristics of states and actions in different tasks. A Multi-Layer Perceptron (MLP) is sufficient for most DRL tasks due to its ability to model complex reward functions, while specialized architectures like U-Net can be advantageous for tasks requiring fine-grained spatial modeling, such as those involving high-dimensional state spaces (e.g., signal spectrum analysis). 
For an MLP implementation, the computational complexity is $\mathcal{O}\left( K(h_L (h_D)^2 + h_D I_D)\right) $ for generation and $\mathcal{O}\left( h_b h_L (h_D)^2 + h_b h_L h_D\right) $ for training, where $K$ denotes the number of denoising steps, $h_L$ is the number of hidden layers, $h_D$ represents the hidden dimension, $I_D$ is the input dimension, and $h_b$ is the batch size, with the ${h_D}^2$ term arising from matrix multiplication in each hidden layer and $h_D I_D$ from input layer computations.

{\textbf{Training Process and Replay Buffers}:} 
Two separate replay buffers, $\mathcal{D}_P$ and $\mathcal{D}_R$, are initialized for the DRL and DRESS training processes, respectively. 
Without loss of generality, we consider a policy-based DRL architecture, specifically the SAC~\cite{haarnoja2018soft}, which has demonstrated strong performance in continuous control tasks. 
During interactions with the environment, the policy agent collects transitions in the form $\{{\bm{s}}_t, {\bm{a}}_t, {\bm{s}}_{t+1}, r^{(E)}_t\}$ in $\mathcal{D}_P$, while the DRESS agent stores transitions as $\{{\bm{s}}_t^G, r^{(G)}, {\bm{s}}_{t+1}^{(G)}, r^{(E)}_t\}$ in $\mathcal{D}_R$. 
We can employ an off-policy approach to train both the policy agent and the DRESS agents simultaneously. During the model update phase, these agents sample mini-batches of transitions from their respective replay buffers and compute their losses according to \eqref{actortrain} and \eqref{dressedsac}. 
The auxiliary reward signal $r^{(G)}_t$ is generated by the most recently updated diffusion-based reward model through the denoising process defined in~\eqref{denoise2} and \eqref{denoise3}. The policy agent and DRESS generation agent are optimized independently using their respective RL methods, enabling modular adaptability and facilitating easy integration with various DRL frameworks.

\begin{algorithm}[t]
\caption{DRESSed-DRL}
\label{alg:simplified_dressed_drl}
\hspace*{0.02in} {\bf Input:}
\begin{itemize}
\item Parameters: DRL algorithm $\cal A$, denoising timesteps $K$, factor $\beta \in [0, 1]$, {\textit{generation agent}} in DRESS with network parameters $\theta$ and $\phi$, {\textit{evaluation agent}} in DRESS with network parameters $\upsilon$.
\item Environment
\end{itemize}
\hspace*{0.02in} {\bf Output:}
\begin{itemize}
\item Trained DRESSed-DRL $\cal A'$
\item Generative and evaluation agents in DRESS
\end{itemize}
\begin{algorithmic}[1]
\State Initialize target networks and replay buffer $\mathcal{D}_R$
\For{each training iteration}
\State \textit{\# Collect experience:}
\State Reset environment and observe initial state $s_0$
\While{not episode terminated}
\State Sample action according to DRL algorithm $\cal A$, e.g., $a_t \sim \pi(\cdot|s_t)$
\State Execute $a_t$, and observe the next state $s_{t+1}$, environment reward $r_t^{(E)}$, and termination indicator $d_t \in \{0,1\}$

\State \textit{\# Reward shaping process:}
\State Generate the reward signal using \eqref{denoise2} and \eqref{denoise3}.
\State Compute augmented reward using~\eqref{combinere}
\State Store transition $(s_t, a_t, r_G, s_{t+1}, r_E, d_t)$ in $\mathcal{D}_R$

\State $s_t \leftarrow s_{t+1}$, $t \leftarrow t + 1$

\If{$d_t = 1$}
\State Break
\EndIf
\EndWhile

\State \textit{\# Update phase:}
\State Sample mini-batch of transitions from $\mathcal{D}_R$
\State Optimize {\textit{generation agent}} using \eqref{actortrain}
\State Optimize {\textit{evaluation agent}} using \eqref{qualitytrain}
\State Update DRL algorithm $\cal A$ using augmented rewards $\mapsto$ DRESSed-DRL $\cal A'$
\EndFor
\end{algorithmic}
\end{algorithm}




\section{Experimental Results}\label{secnums}
The central focus of this paper is to develop a universal reward generation framework, i.e., DRESS, that can enhance DRL performance in wireless network optimization tasks with sparse reward signals. Thus, our experiments are structured to investigate the following questions:

\begin{enumerate}
\item[{\textbf{Q1)}}] \textbf{Effectiveness:} How does DRESS improve DRL performance in wireless environments with sparse rewards compared to baseline approaches? 

\item[{\textbf{Q2)}}] \textbf{Robustness:} How sensitive is DRESS to different hyperparameters, environmental conditions, and used DRL architectures? 

\item[{\textbf{Q3)}}]  \textbf{Generalizability:} To what extent can DRESS adapt to different DRL algorithms and optimization scenarios? 
\end{enumerate}
We first describe our experimental setup, followed by comprehensive results addressing these questions. 

\subsection{Experiments Setting}
\begin{figure}[t]
\centering
\includegraphics[width=0.45\textwidth]{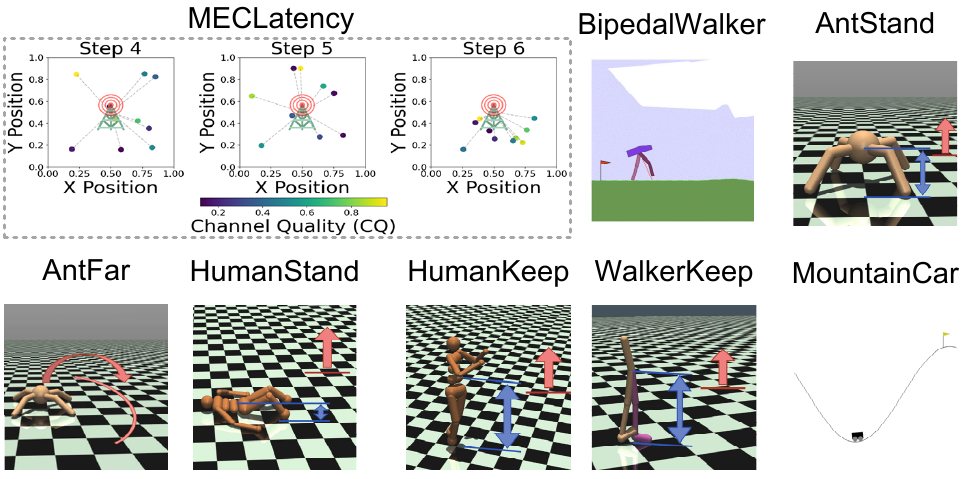}
\vspace{-0.2cm}
\caption{Benchmark environments. {\textit{MECLatency}} for robust wireless network optimization simulation and seven classical control tasks.}
\label{fig:generalenv}
\end{figure}
\subsubsection{Wireless Benchmark Environment} 
We design our customized environment {\textit{MECLatency}} according to the system model proposed in Section.~\ref{environment} using the OpenAI Gym interface~\cite{brockman2016openai}. In the simulations, the key parameters are configured as follows: the learning rate is set to 200, the maximum bandwidth is 20 Hz, the maximum power is 10 W, and the maximum computation capacity is 100. The users are uniformly and randomly distributed within a circular area with a diameter of 10 km, with the number of users ranging from 0 to 10, traveling at low to moderate speeds (e.g., no faster than 60 km/h)~\cite{R5}. In this setting, the association between the user and the BS remains relatively stable before the resource allocation results are returned to the users.

\subsubsection{DRL Benchmark Environments}  As shown in Fig.~\ref{fig:generalenv}, we have conducted simulations on the proposed {\textit{MECLatency}} environment and other environments based on the OpenAI Gym interface~\cite{brockman2016openai} handling continuous control tasks with
challenging sparse and delayed rewards. Besides {\textit{MECLatency}}, the environments used to verify the generalizablity of our algorithm include: (1) {\textit{BipedalWalker}}: train a bipedal robot to walk across rough terrain while maintaining balance and moving forward; (2) {\textit{AntStand}}: teach a quadrupedal (ant-like) robot to stand up from a lying position; (3) {\textit{AntFar}}: train a quadrupedal (ant-like) robot to walk across complex terrain and travel longer distances; (4) {\textit{HumanStand}}: teach a humanoid robot to stand up and maintain stability; (5) {\textit{HumanKeep}}: train a humanoid robot to remain standing despite external disturbances; (6) {\textit{WalkerKeep}}: enable a bipedal robot to continue walking in a dynamic environment without falling; (7) {\textit{MountainCar}}: train a small car to use momentum to climb a steep hill and reach the goal. These environments cover different domains of {\textit{MuJoCo}}~\cite{todorov2012mujoco}, arm robot~\cite{de2023gymnasium}, and physical control~\cite{towers2024gym}. These environments are challenging simulations for testing RL algorithms, featuring complex nonlinear dynamics, continuous action spaces, and high-dimensional state spaces.

\subsubsection{Benchmark Algorithms} We compare our approach with several reward shaping-based algorithms and state-of-the-art DRL methods. The reward shaping algorithms include: (1) the advanced reward agent-based method from~\cite{mareward}, which uses a policy agent for optimal behavior learning while a reward agent generates auxiliary reward signals; (2) RL Optimizing Shaping Algorithm (ROSA)~\cite{mguni2023learning}, which implements reward shaping through a two-player Markov game with an auxiliary agent; and Exploration Guided Reward Shaping (ExploRS) algorithm~\cite{devidze2022exploration}, which combines an intrinsic reward function with exploration-based bonuses. For DRL baseline comparison, we also evaluate against standard DRL algorithms without reward shaping, including SAC, Twin Delayed Deep Deterministic Policy
Gradient (TD3)~\cite{fujimoto2018addressing}, Deep Deterministic Policy Gradient (DDPG)~\cite{lillicrap2015continuous}, and REINFORCE~\cite{bhatnagar2023reinforce}. 

The experimental platform is built on a workstation with Ubuntu 20.04
, powered by an AMD EPYC 7543 32-Core CPU and 4 NVIDIA RTX 4060 GPUs. For most conventional network optimization problems, mainstream consumer-grade CPU can support DRESS. For DRL benchmark testing, we recommend at least one NVIDIA 30-series GPU or equivalent.

\subsection{Experiments Performance Analysis}
\subsubsection{For Q1-Effectiveness}
\begin{figure*}[t]
\centering
\subfigure[Latency limit is $0.02$ and $\beta$ is $0.2$.]
{
\includegraphics[width=0.23\textwidth, height=0.18\textwidth]{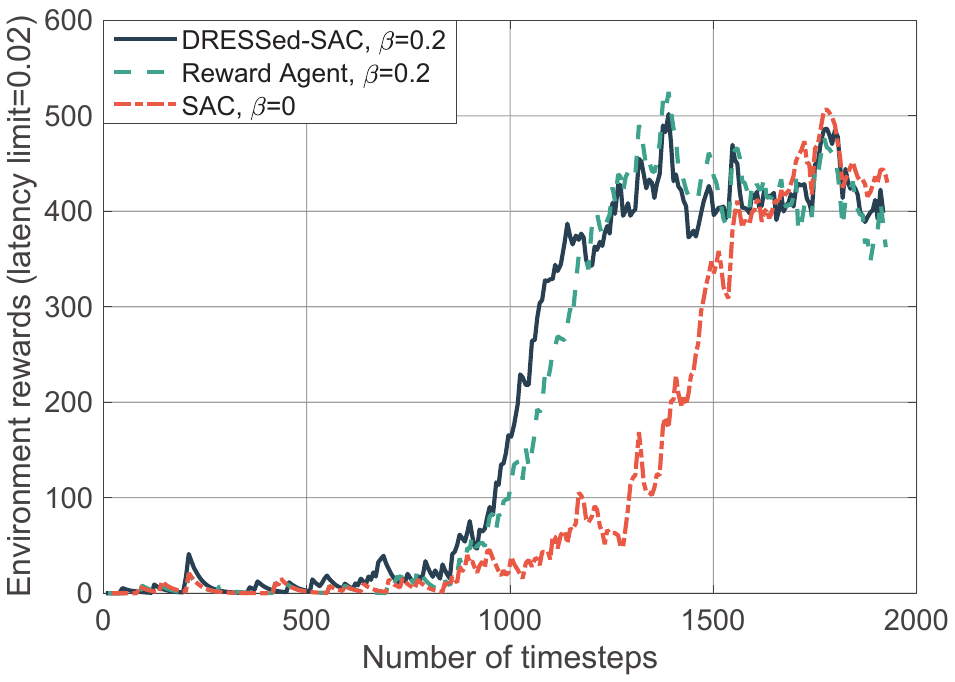}
}
\subfigure[Latency limit is $0.02$ and $\beta$ is $0.4$.]
{
\includegraphics[width=0.23\textwidth, height=0.18\textwidth]{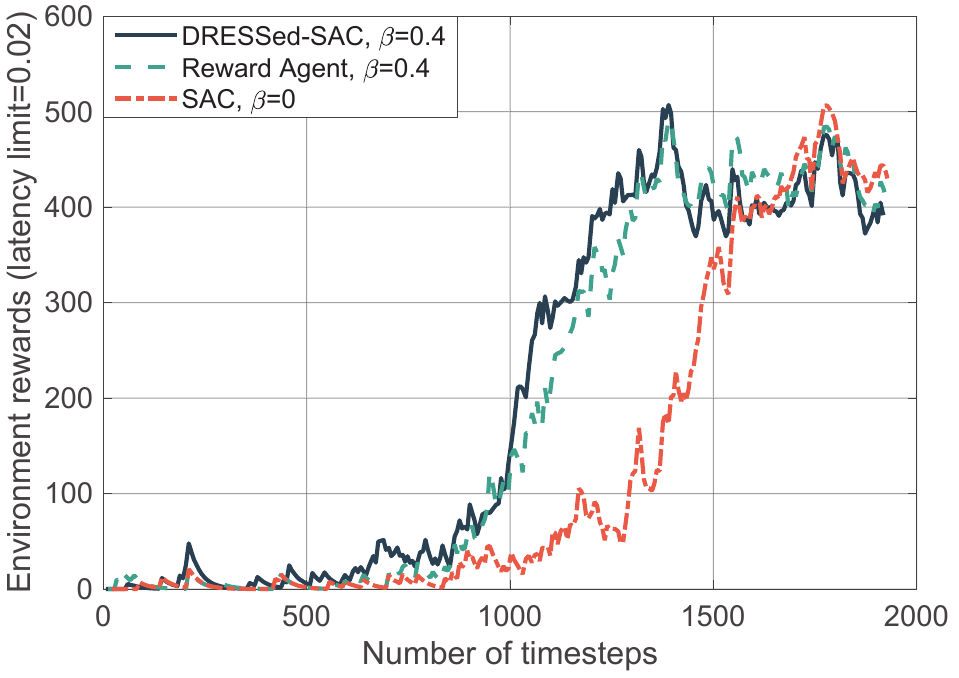}
}
\subfigure[Latency limit is $0.02$ and $\beta$ is $0.6$.]
{
\includegraphics[width=0.23\textwidth, height=0.18\textwidth]{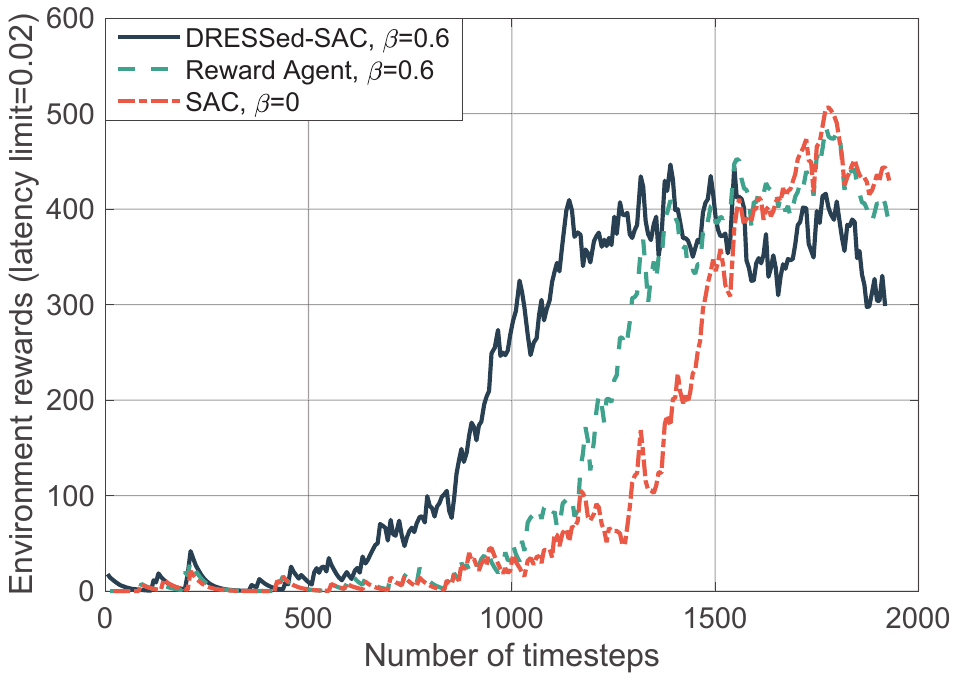}
}
\subfigure[Latency limit is $0.02$ and $\beta$ is $0.8$.]
{
\includegraphics[width=0.23\textwidth, height=0.18\textwidth]{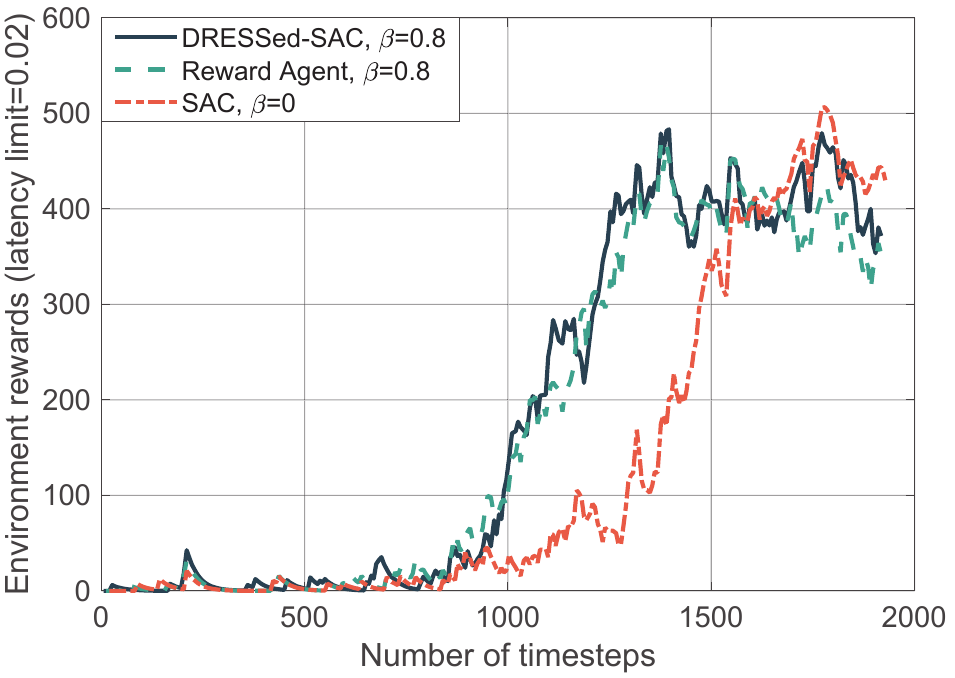}
}
\caption{Comparison of training performance in {\textit{MECLatency}} with a latency limit of $0.02$ and varying $\beta$ values (0.2, 0.4, 0.6, and 0.8) across DRESSed-SAC, reward agent-based method, and standard SAC.}
\label{fig:latency_0.02}
\end{figure*}
\begin{figure*}[t]
\centering
\subfigure[Latency limit is $0.02$ and $\beta$ is $0.2$.]
{
\includegraphics[width=0.23\textwidth, height=0.18\textwidth]{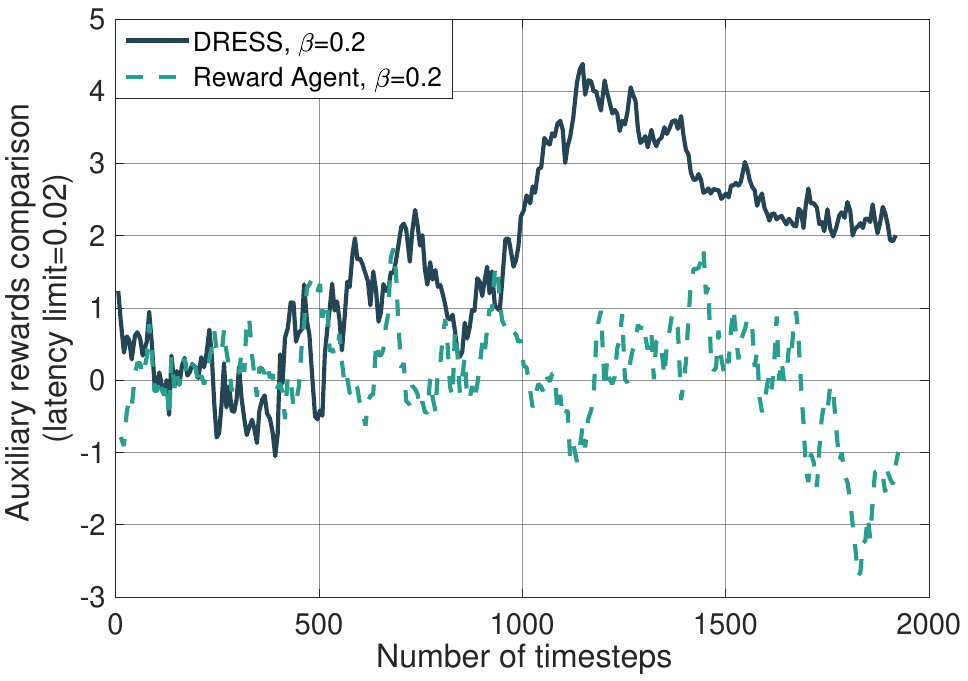}
}
\subfigure[Latency limit is $0.02$ and $\beta$ is $0.4$.]
{
\includegraphics[width=0.23\textwidth, height=0.18\textwidth]{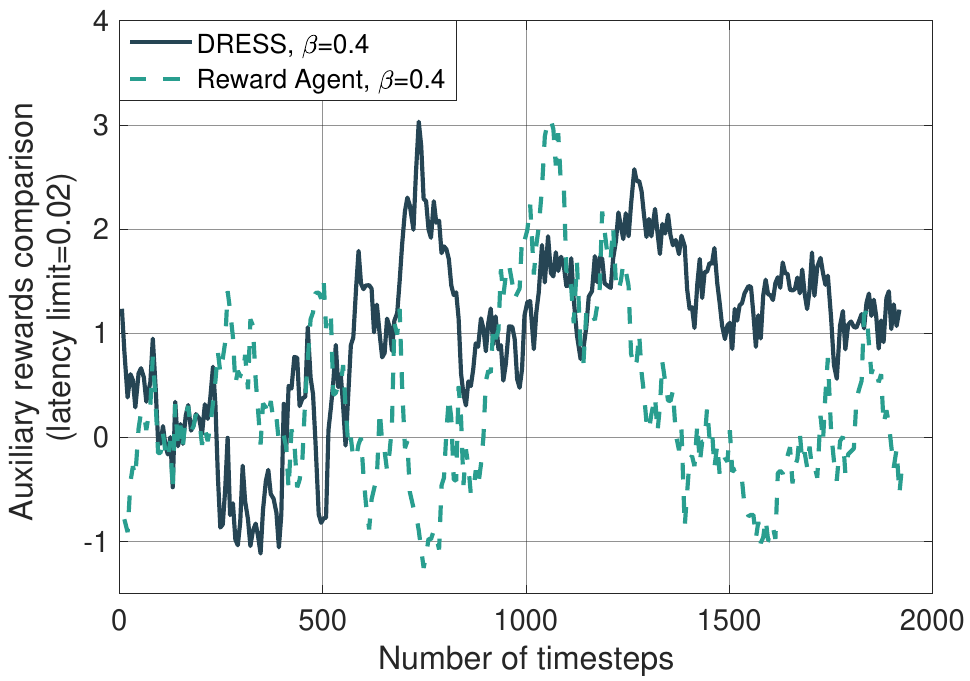}
}
\subfigure[Latency limit is $0.02$ and $\beta$ is $0.6$.]
{
\includegraphics[width=0.23\textwidth, height=0.18\textwidth]{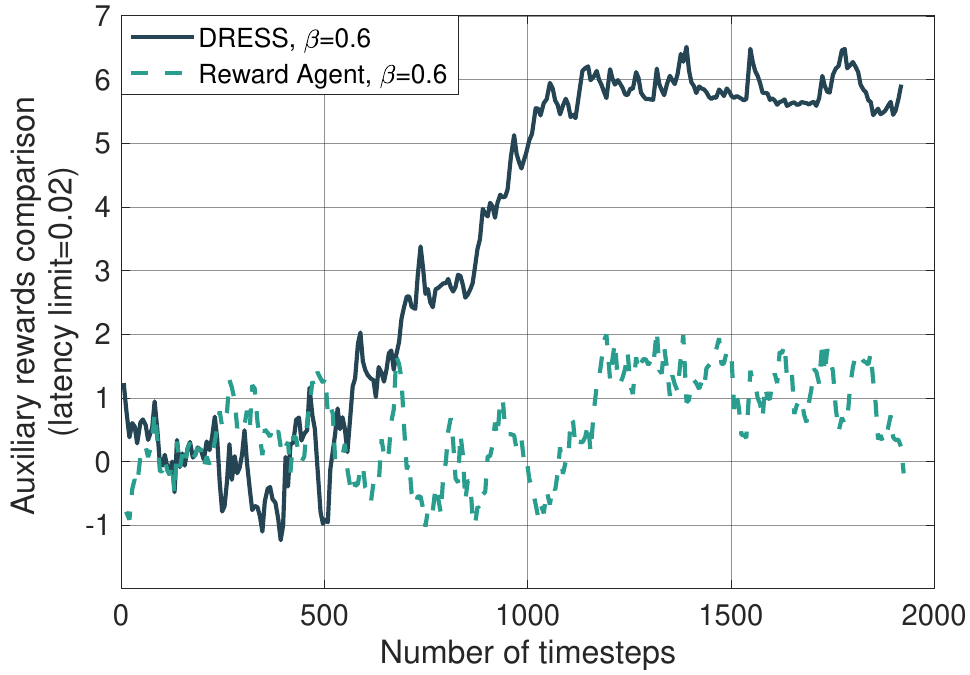}
}
\subfigure[Latency limit is $0.02$ and $\beta$ is $0.8$.]
{
\includegraphics[width=0.23\textwidth, height=0.18\textwidth]{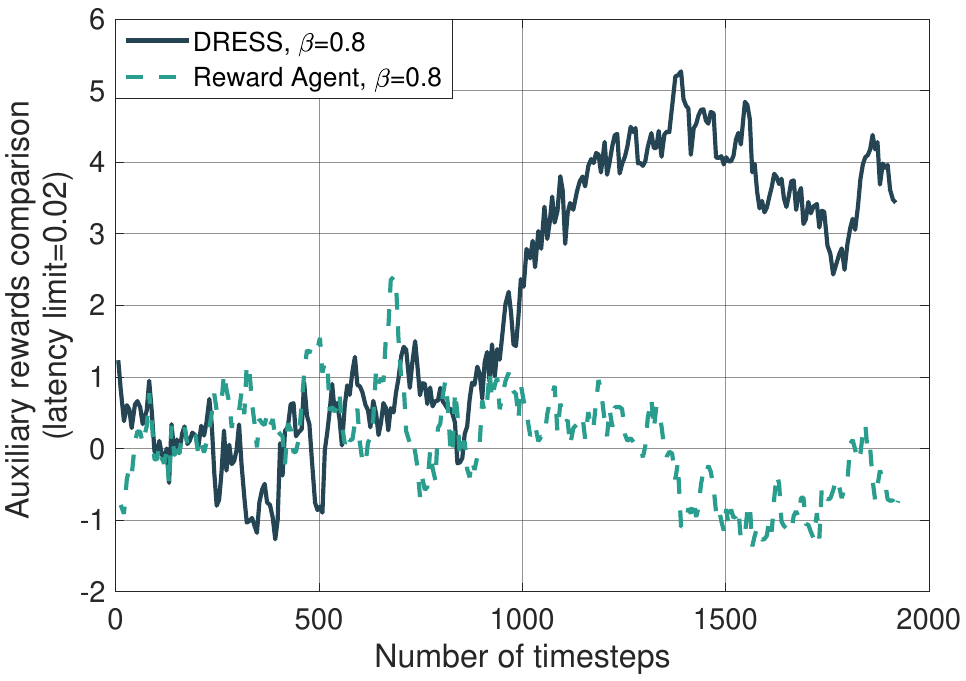}
}
\caption{Comparison of auxiliary reward signals during the training process in {\textit{MECLatency}} with a latency limit of $0.02$ and varying $\beta$ values ($0.2$, $0.4$, $0.6$, and $0.8$) across DRESSed-SAC and reward agent-based method.}
\label{ra_0.02}
\end{figure*}

\begin{figure*}[t]
\centering
\subfigure[Latency limit is $0.01$ and $\beta$ is $0.2$.]
{
\includegraphics[width=0.23\textwidth, height=0.18\textwidth]{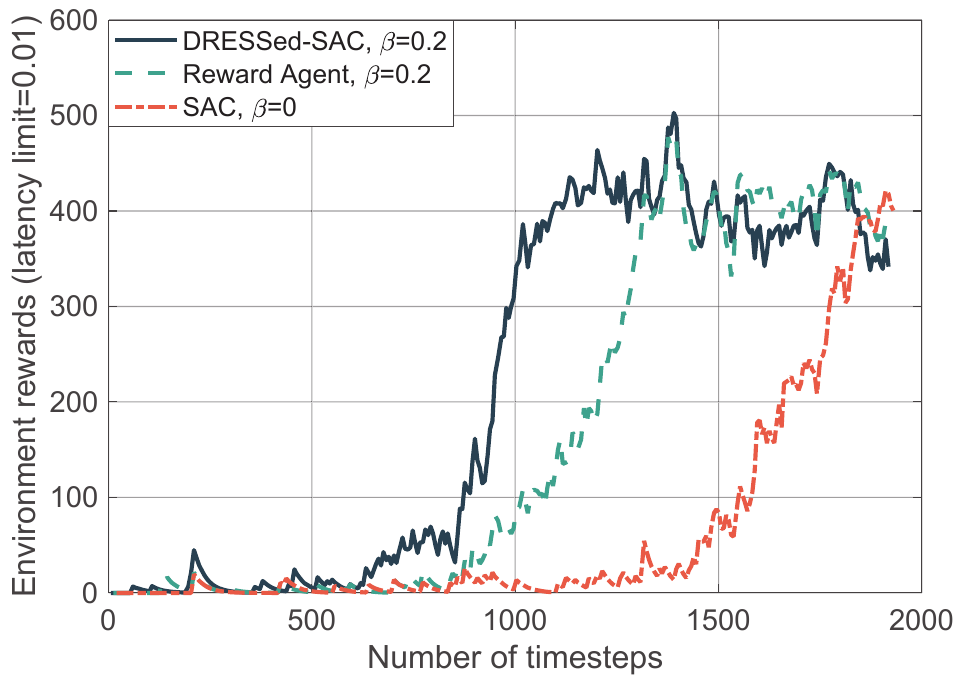}
}
\subfigure[Latency limit is $0.01$ and $\beta$ is $0.4$.]
{
\includegraphics[width=0.23\textwidth, height=0.18\textwidth]{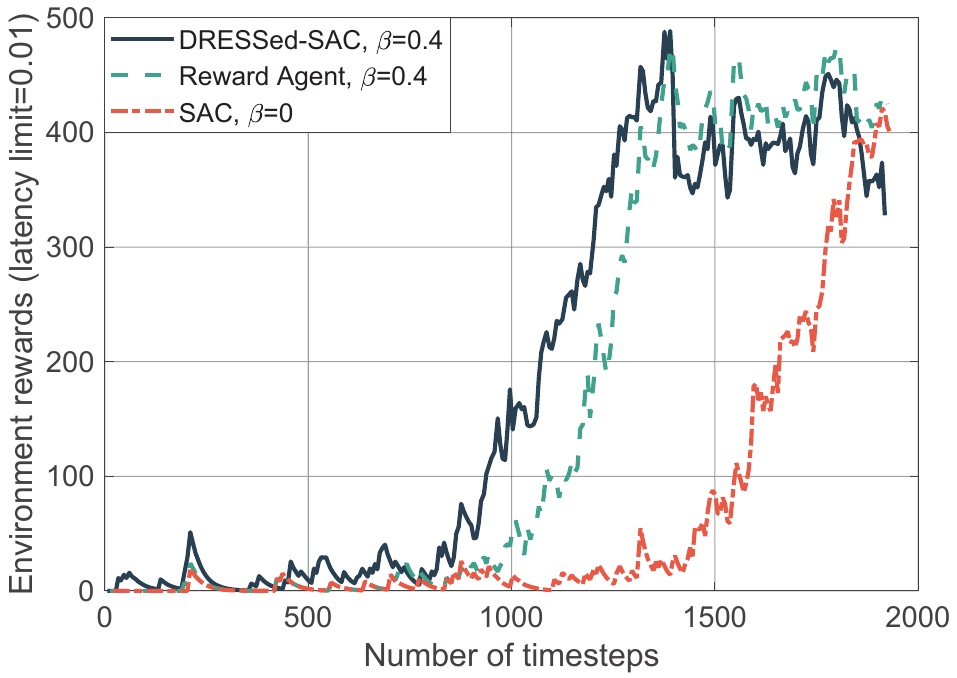}
}
\subfigure[Latency limit is $0.01$ and $\beta$ is $0.6$.]
{
\includegraphics[width=0.23\textwidth, height=0.18\textwidth]{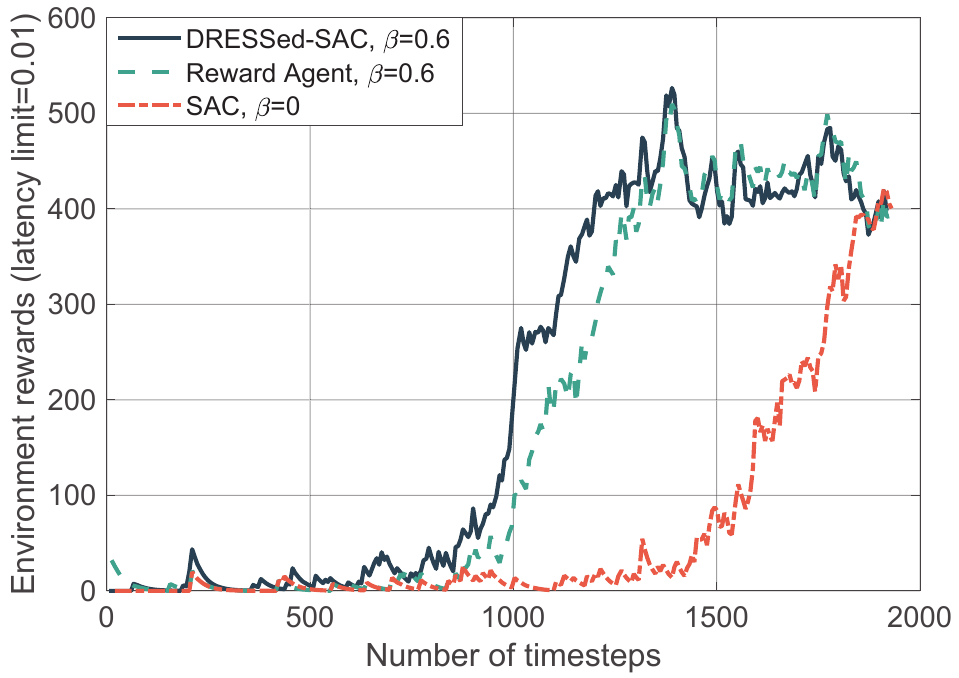}
}
\subfigure[Latency limit is $0.01$ and $\beta$ is $0.8$.]
{
\includegraphics[width=0.23\textwidth, height=0.18\textwidth]{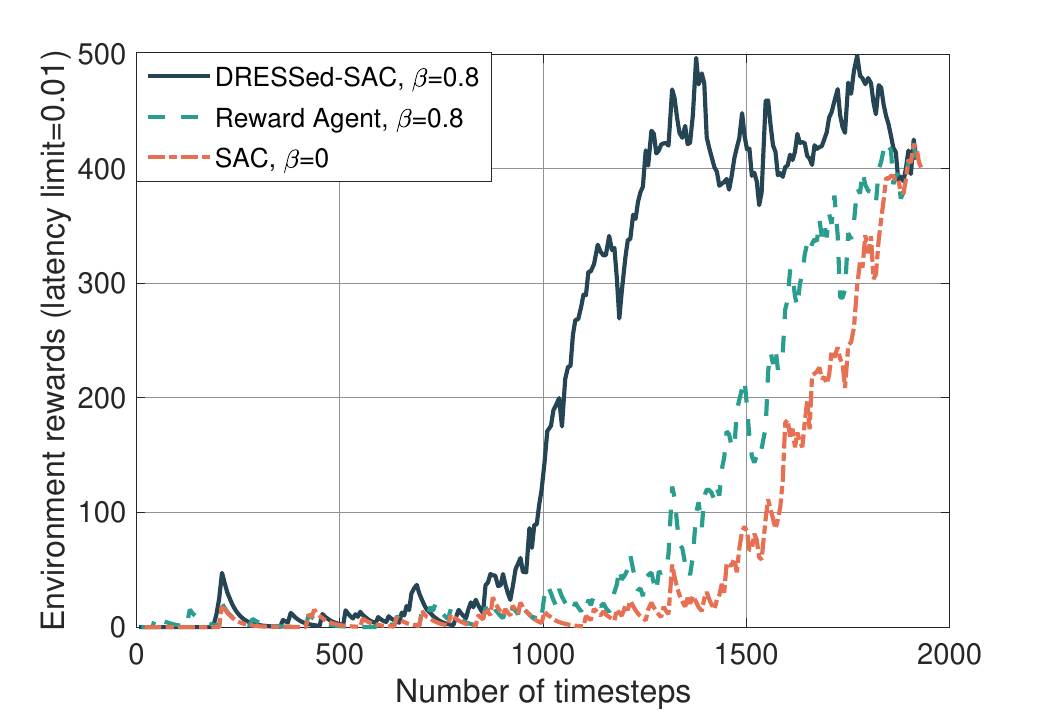}
}
\caption{Comparison of training performance in {\textit{MECLatency}} with a latency limit of $0.01$ and varying $\beta$ values ($0.2$, $0.4$, $0.6$, and $0.8$) across DRESSed-SAC, reward agent-based method, and standard SAC.}
\label{fig:latency_0.01}
\end{figure*}

\begin{figure*}[t]
\centering
\subfigure[Latency limit is $0.01$ and $\beta$ is $0.2$.]
{
\includegraphics[width=0.23\textwidth, height=0.18\textwidth]{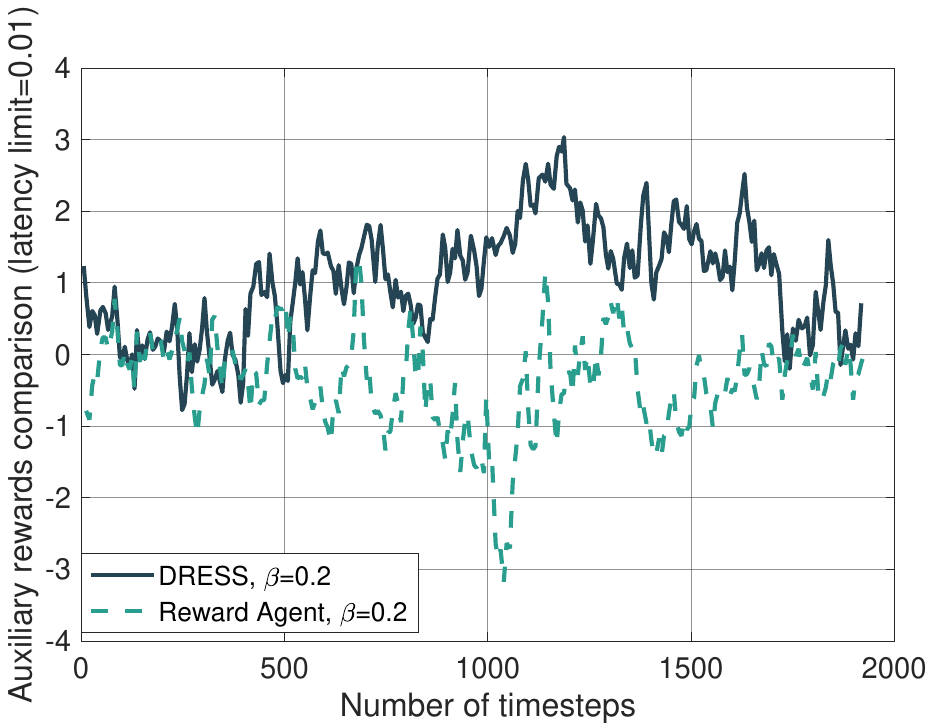}
}
\subfigure[Latency limit is $0.01$ and $\beta$ is $0.4$.]
{
\includegraphics[width=0.23\textwidth, height=0.18\textwidth]{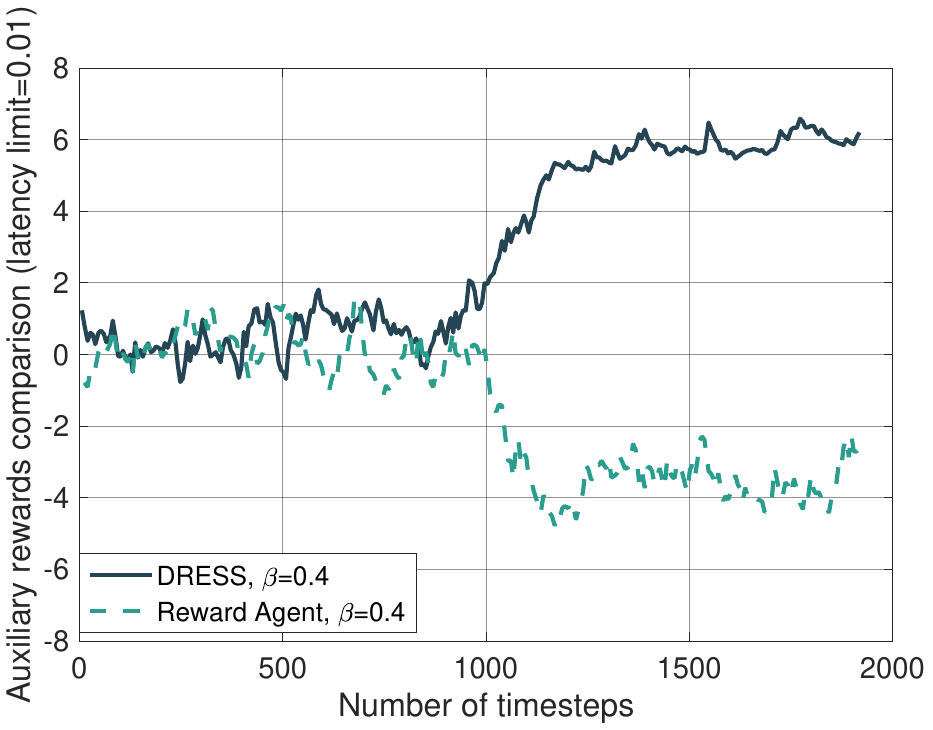}
}
\subfigure[Latency limit is $0.01$ and $\beta$ is $0.6$.]
{
\includegraphics[width=0.23\textwidth, height=0.18\textwidth]{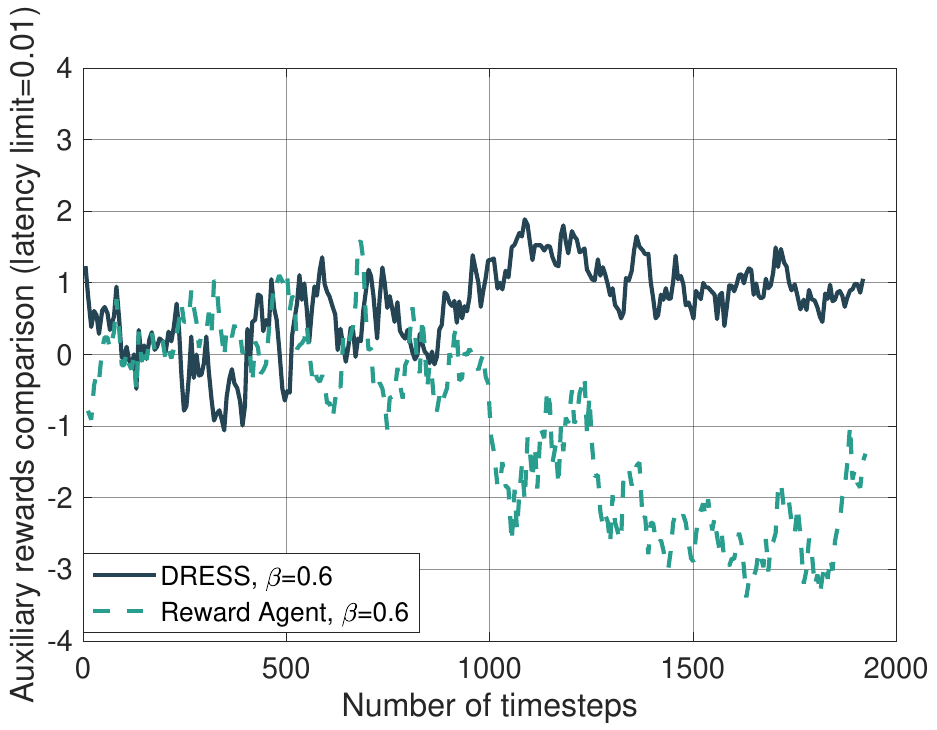}
}
\subfigure[Latency limit is $0.01$ and $\beta$ is $0.8$.]
{
\includegraphics[width=0.23\textwidth, height=0.18\textwidth]{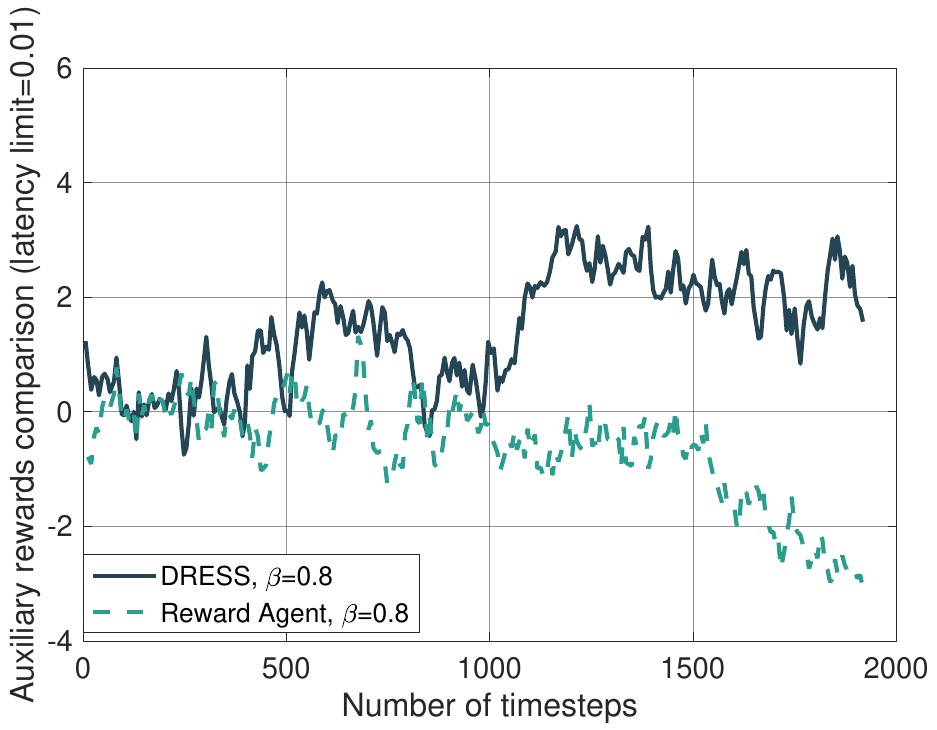}
}
\caption{Comparison of auxiliary reward signals during the training process in {\textit{MECLatency}} with a latency limit of $0.01$ and varying $\beta$ values ($0.2$, $0.4$, $0.6$, and $0.8$) across DRESSed-SAC and reward agent-based method.}
\label{ra_0.01}
\end{figure*}
The effectiveness of DRESS is demonstrated by its ability to enhance the performance of DRL algorithms that previously struggled to converge or converged slowly in extreme network environments. We compare our approach with standard SAC and the state-of-the-art reward agent-based method~\cite{mareward}. 
Fig.~\ref{fig:latency_0.02} illustrates the reward performance of the proposed DRESSed-SAC method compared to the baseline approaches under a latency limit of 0.02. 
The results demonstrate that simply augmenting standard SAC with DRESS immediately yields significant convergence improvements, achieving approximately $1.5{\rm{x}}$ times faster convergence than the original SAC algorithm and being comparable to the carefully designed reward agent-based method. 
Moreover, for reward shaping algorithms, the parameter $\beta$ controls the degree to which auxiliary rewards are added to environmental rewards. Our evaluation across different $\beta$ values further demonstrates the effectiveness of DRESSed-SAC across various configurations. For instance, at $\beta=0.6$, DRESSed-SAC significantly outperforms the reward agent-based method in terms of convergence speed.
As shown in Fig.~\ref{ra_0.02}, the auxiliary reward signals generated by DRESS also exhibit superior characteristics compared to the reward agent method. An ideal auxiliary signal should facilitate exploration in early stages and then stabilize to form a balance with the environmental reward to promote exploitation. The DRESS-generated signals demonstrate this pattern, effectively guiding the agent during early exploration phases while providing more stable signals during later exploitation stages.

To further test whether the algorithm can maintain effectiveness in more extreme environments, we reduced the latency requirement in {\textit{MECLatency}} by half. With a latency requirement of $0.01$, AI algorithms face greater difficulty in obtaining effective rewards. Fig.~\ref{fig:latency_0.01} illustrates the comparison of training performance under this stricter latency limit with varying $\beta$ values. We can observe that under these challenging conditions, the standard SAC algorithm's training curve only begins to rise after $1500$ timesteps. In contrast, DRESSed-SAC is the only approach that has already reached stability by this point. Across various $\beta$ values, DRESSed-SAC demonstrates a more significant improvement in convergence speed compared to the reward agent-based method. Additionally, Fig.~\ref{ra_0.01} shows that the auxiliary rewards from DRESSed-SAC maintain their desirable properties.

\subsubsection{For Q2-Robustness}
\begin{figure}[t]
\centering
\subfigure[Environment reward.]
{
\includegraphics[width=0.22\textwidth]{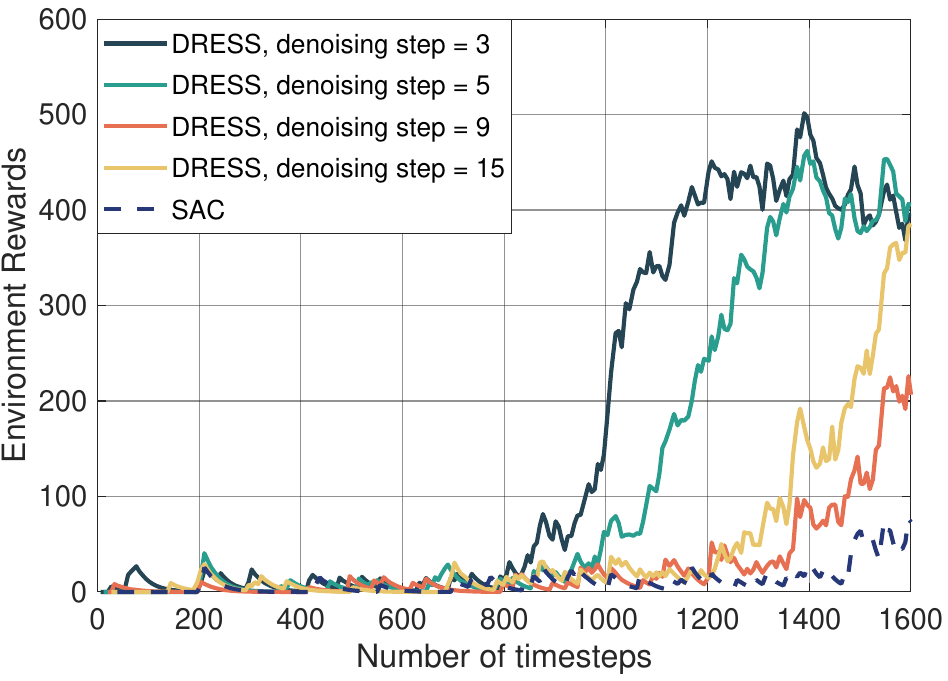}
}
\subfigure[Auxiliary reward.]
{
\includegraphics[width=0.22\textwidth]{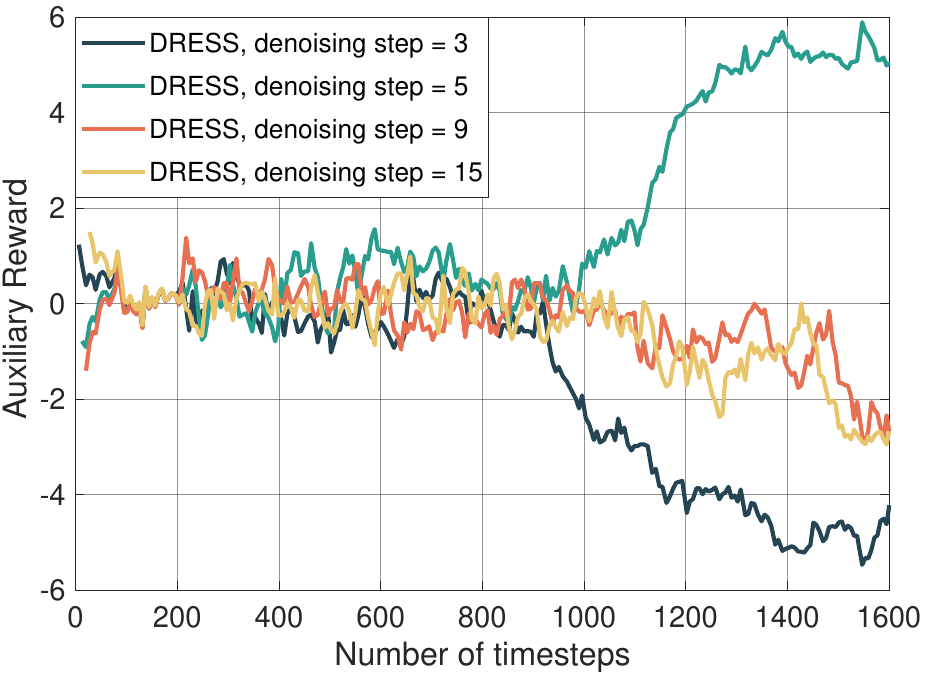}
}
\caption{Environment and auxiliary reward signals of DRESSed-SAC in {\textit{MECLatency}} with a latency limit of $0.01$ and varying denoising step values ($3$, $5$, $9$, and $15$).}
\label{diffu}
\end{figure}

\begin{figure*}[t]
\centering
\subfigure[DRESSed-SAC and SAC.]
{
\includegraphics[width=0.48\textwidth]{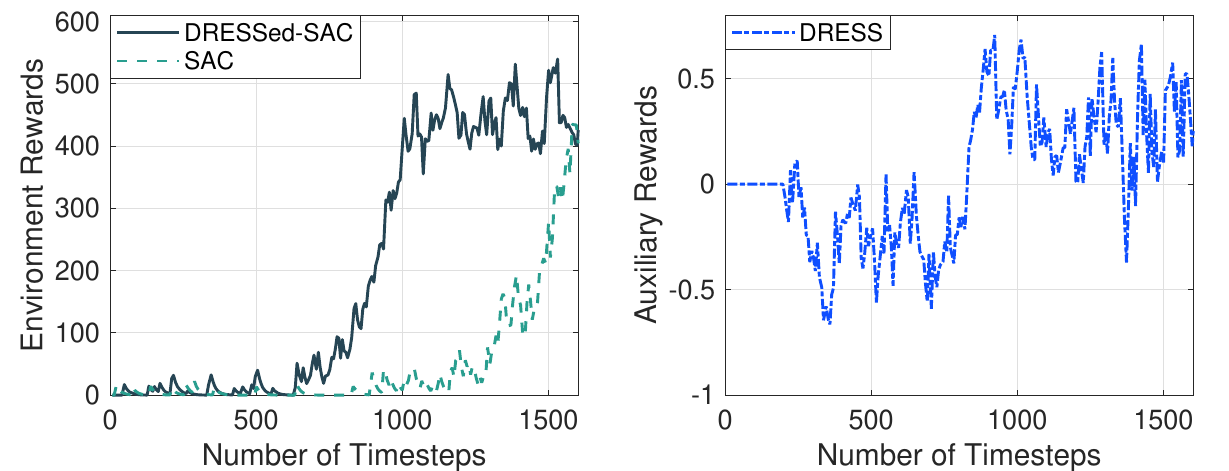}
}
\subfigure[DRESSed-TD3 and TD3.]
{
\includegraphics[width=0.48\textwidth]{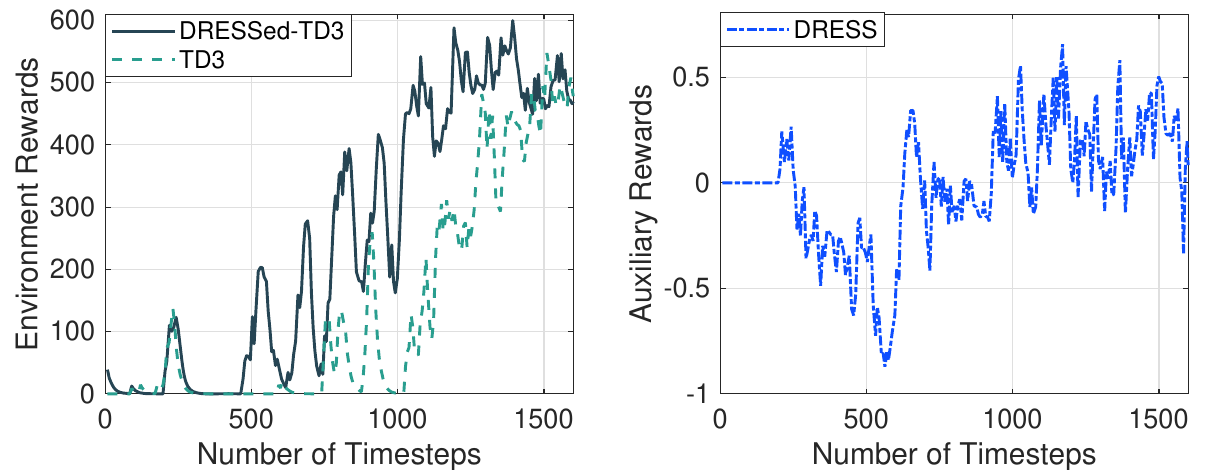}
}
\subfigure[DRESSed-DDPG and DDPG.]
{
\includegraphics[width=0.48\textwidth]{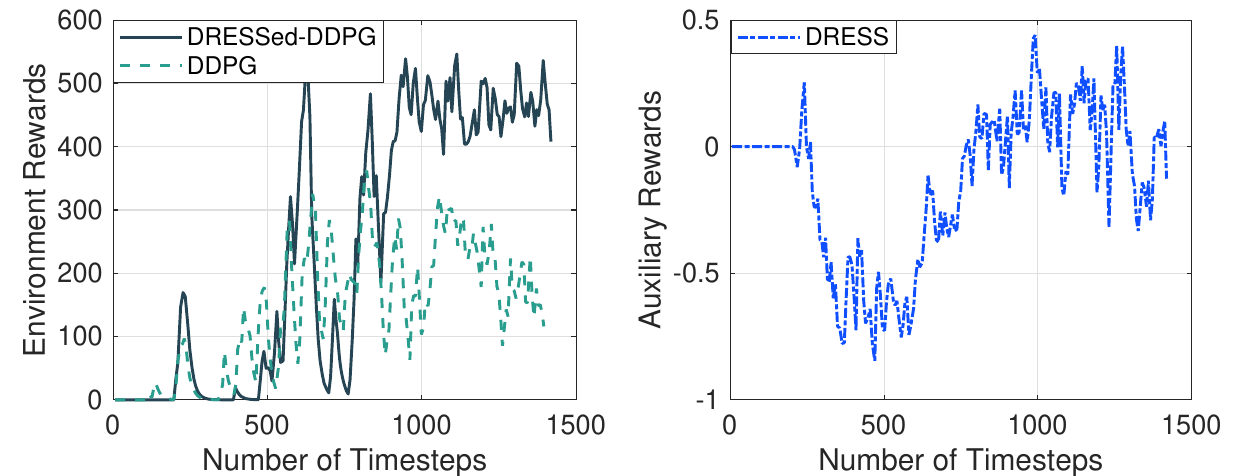}
}
\subfigure[DRESSed-REINFORCE and REINFORCE.]
{
\includegraphics[width=0.47\textwidth]{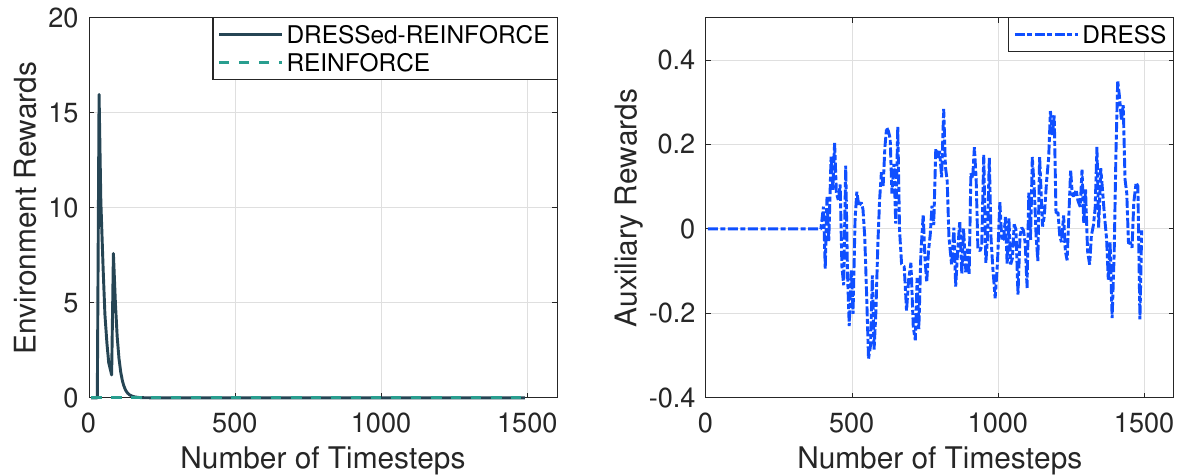}
}
\caption{Environment and auxiliary rewards of DRESSed-DRL and DRL algorithms in {\textit{MECLatency}} with a latency limit of $0.02$ and $\beta = 0.2$.}
\label{fig:comparedrl}
\end{figure*}
The robustness of DRESS has three key aspects: hyperparameter robustness, DRL architecture robustness, and environmental robustness. For hyperparameter robustness, our analysis of effectiveness already demonstrated the robustness to the $\beta$. Another critical hyperparameter in DRESS is the number of denoising steps used to generate auxiliary signals. Fig.~\ref{diffu} illustrates the environment and auxiliary reward signals of DRESSed-SAC in {\textit{MECLatency}} with a latency limit of $0.01$ and varying denoising step values ($3$, $5$, $9$, and $15$). 
We observe that regardless of the step value, DRESSed-SAC consistently shows improved convergence speed compared to standard SAC. The shape of the auxiliary reward signals varies with different step settings, but all demonstrate a transition from exploration to stability, representing a balance with the environmental rewards.
A deeper insight into the denoising step parameter $K$'s effect reveals the \textit{exploration-exploitation balance:} DRESS offers flexibility in balancing exploration and exploitation by adjusting $K$. A smaller $K$ introduces diversity in the generated rewards that encourage exploratory behaviors, while a larger $K$ enhances state-action conditioning through the generation process, yielding relatively deterministic reward signals that stabilize policy exploitation.

The second aspect concerns robustness across different DRL architectures. This architectural robustness demonstrates DRESS's versatility as a universal reward shaping framework that can enhance various DRL algorithms without requiring structural modifications. Fig.~\ref{fig:comparedrl} illustrates the environment and auxiliary rewards of DRESSed-DRL variants compared to their standard counterparts in \textit{MECLatency} with a latency limit of $0.02$ and $\beta = 0.2$. We showcase four DRL algorithms with and without DRESS integration: SAC, TD3, DDPG, and REINFORCE. By the 1500-timestep mark, with the exception of REINFORCE, all algorithms demonstrate significant performance improvements when augmented with DRESS. For high-performing algorithms like SAC, DRESS accelerates convergence speed, as shown in Fig.~\ref{fig:comparedrl} (a). For algorithms like DDPG that struggle to converge effectively, DRESS doubles the final environmental reward values achieved, as shown in Fig.~\ref{fig:comparedrl} (c). The REINFORCE algorithm represents a special case where both the original and DRESSed versions achieve near-zero rewards at 1500 timesteps, likely due to its policy gradient approach requiring more samples in complex continuous control environments like \textit{MECLatency}.

The third aspect is environmental robustness. DRESS maintains performance across parameter variations within the \textit{MECLatency} task, as shown by consistent high performance with different latency constraints (Fig.~\ref{fig:latency_0.02} and Fig.~\ref{fig:latency_0.01}). DRESS also performs robustly across different environments, including standard GYM benchmarks. This extends into generalizability, examined in the following discussion.

\subsubsection{For Q3-Generalizablity}
\begin{table*}[t]
\caption{The average episodic returns and standard errors of DRESSed-SAC and the benchmark algorithms.}
\centering
{\small\begin{tabular}{lccccccc}
\hline
Environments & DRESSed-SAC & Reward Agent & ROSA & ExploRS & SAC & TD3 \\
\hline
BipedalWalker & $\mathbf{155.19 \pm 3.36}$ & $76.15 \pm 4.35$ & $30.27 \pm 2.43$ & $5.65 \pm 0.82$ & $68.27 \pm 9.01$ & $0.00 \pm 0.00$  \\
AntStand & $\mathbf{38.44 \pm 0.44}$ & $28.66 \pm 1.82$ & $3.80 \pm 0.03$ & $4.52 \pm 0.04$ & $15.93 \pm 0.69$ & $0.00 \pm 0.00$  \\
AntFar & $\mathbf{80.14 \pm 0.71}$ & $67.77 \pm 4.30$ & $4.71 \pm 0.28$ & $5.42 \pm 0.22$ & $0.64 \pm 0.57$ & $0.81 \pm 0.02$\\
HumanStand & $21.79 \pm 0.22$ & $\mathbf{29.72 \pm 1.85}$ & $8.55 \pm 0.03$ & $8.63 \pm 0.03$ & $9.31 \pm 0.05$ & $5.72 \pm 0.04$  \\
HumanKeep & $\mathbf{182.91 \pm 0.63}$ & $160.31 \pm 7.30$ & $152.38 \pm 4.98$ & $158.09 \pm 4.42$ & $4.59 \pm 0.84$ & $0.00 \pm 0.00$  \\
WalkerKeep  & $\mathbf{144.02 \pm 0.85}$ & $77.14 \pm 8.77$ & $32.14 \pm 1.19$ & $2.47 \pm 0.13$ & $70.96 \pm 8.10$ & $0.00 \pm 0.00$  \\
MountainCar & $\mathbf{94.47 \pm 0.06}$ & $0.89 \pm 0.01$ & $-0.90 \pm 0.02$ & $-0.99 \pm 0.02$ & $-0.05 \pm 0.02$ & $0.00 \pm 0.00$  \\
\hline
\end{tabular}}
\label{tablec}
\end{table*}

\begin{figure}[t]
\centering
\subfigure[Environment reward]
{
\includegraphics[width=0.226\textwidth]{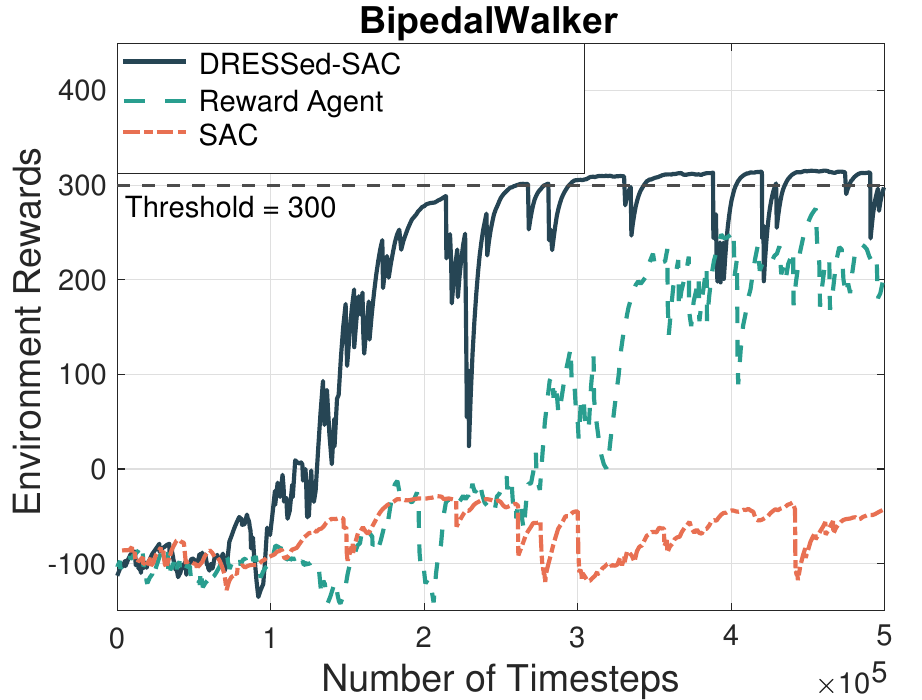}
}
\subfigure[Auxiliary reward]
{
\includegraphics[width=0.226\textwidth]{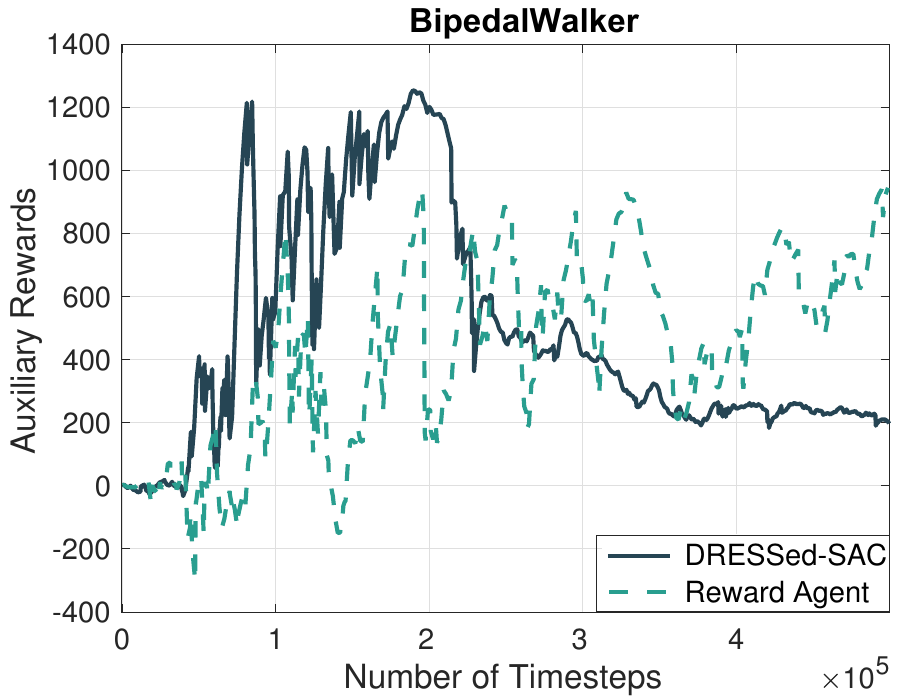}
}
\caption{Comparison of environment and auxiliary rewards in {\textit{BipedalWalker}} across DRESSed-SAC, reward agent-based method, and SAC.}
\label{fig:drlenvs}
\end{figure}

\begin{figure*}[t]
\centering
\subfigure[AntStand environment.]
{
\includegraphics[width=0.3\textwidth]{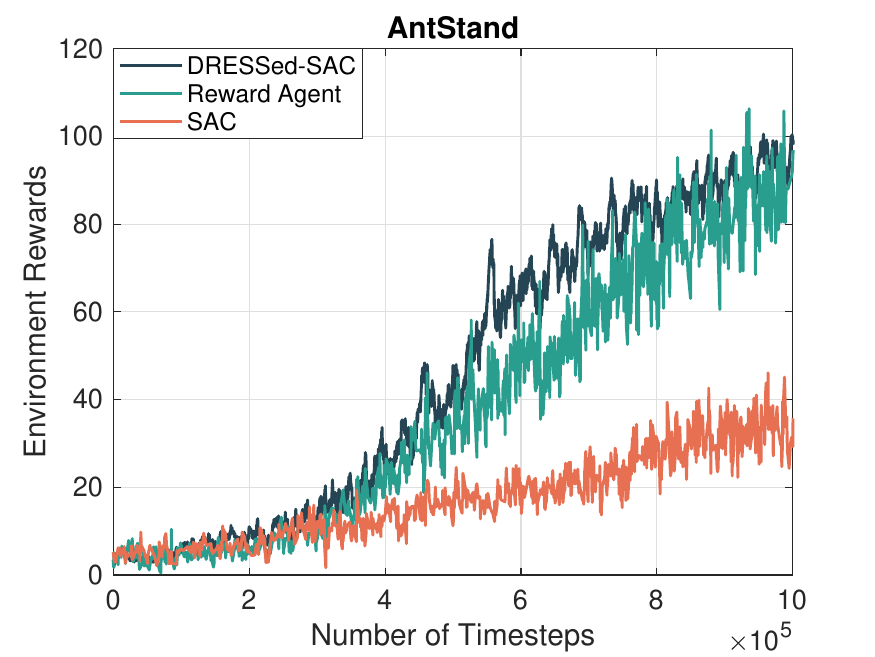}
}
\subfigure[AntFar environment.]
{
\includegraphics[width=0.3\textwidth]{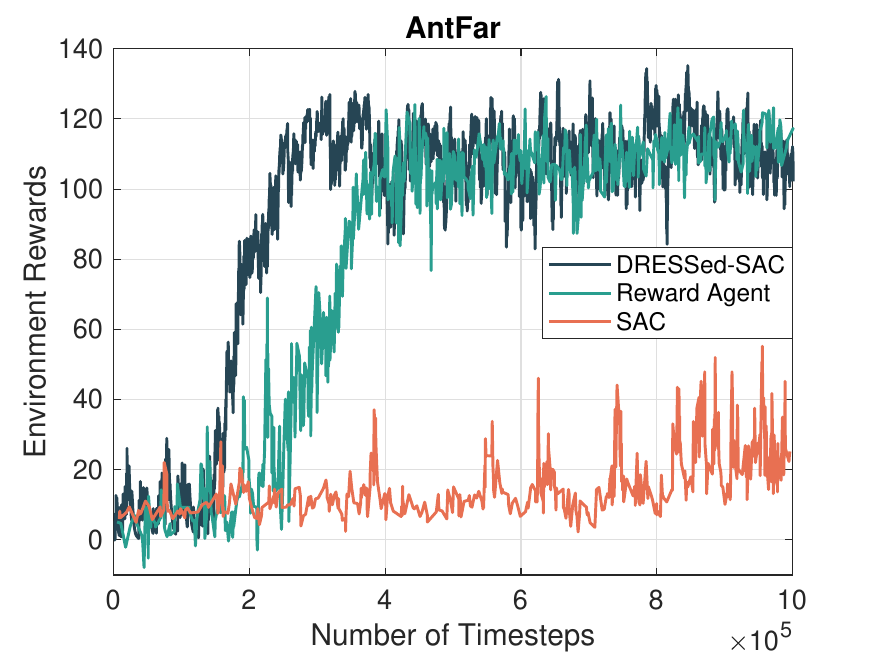}
}
\subfigure[HumanStand environment.]
{
\includegraphics[width=0.3\textwidth]{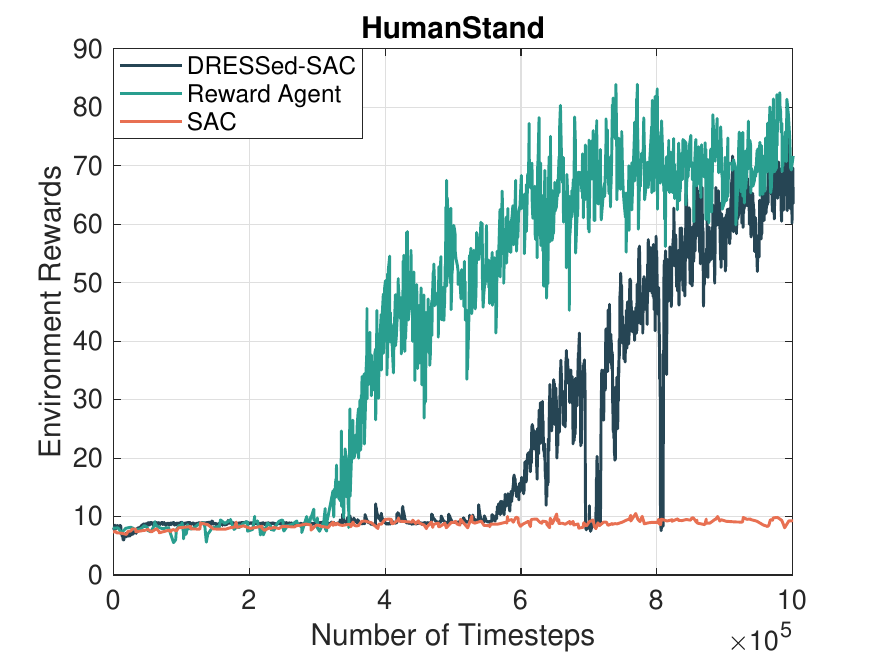}
}
\subfigure[HumanKeep environment.]
{
\includegraphics[width=0.3\textwidth]{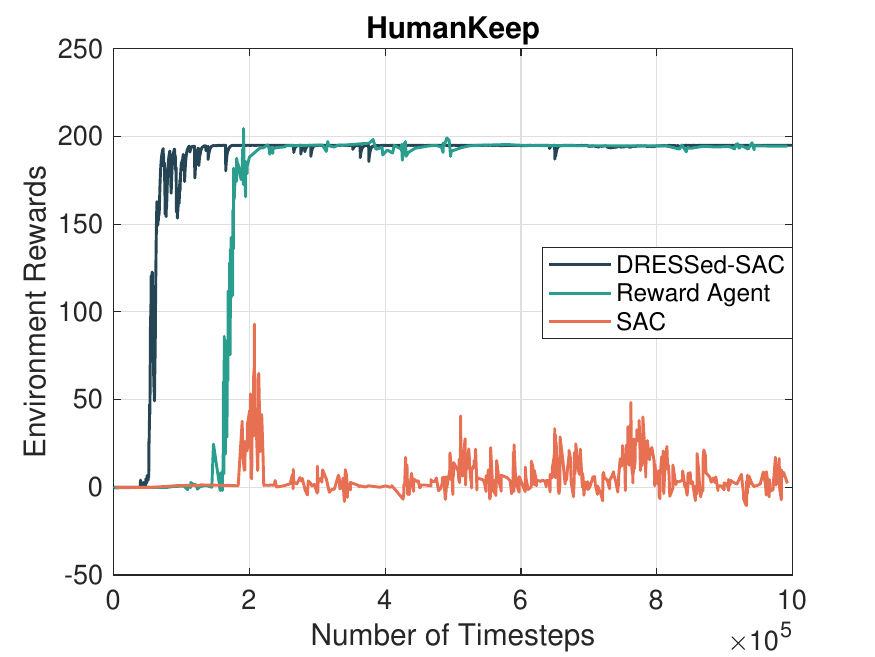}
}
\subfigure[WalkerKeep environment.]
{
\includegraphics[width=0.3\textwidth]{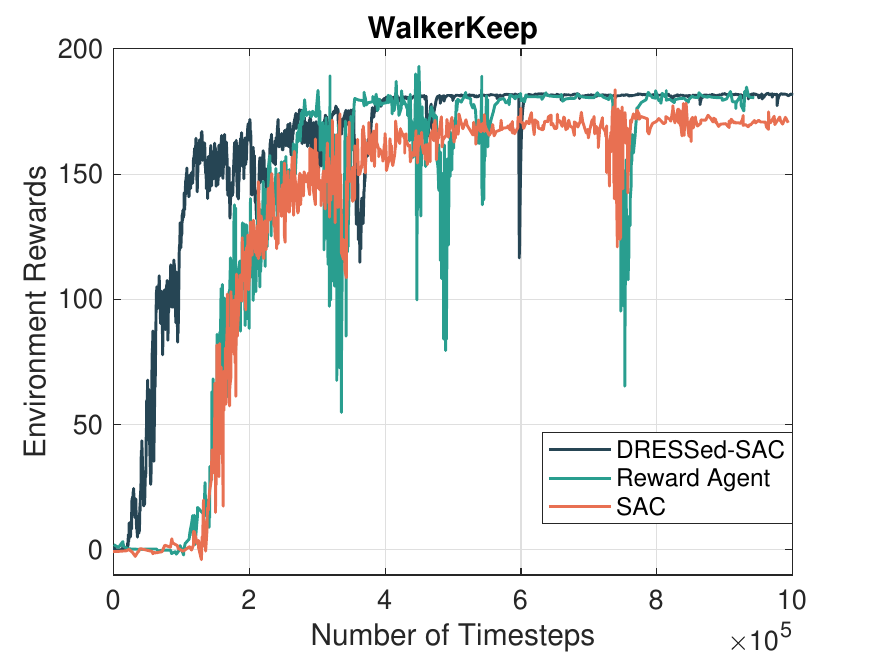}
}
\subfigure[MountainCar environment.]
{
\includegraphics[width=0.3\textwidth]{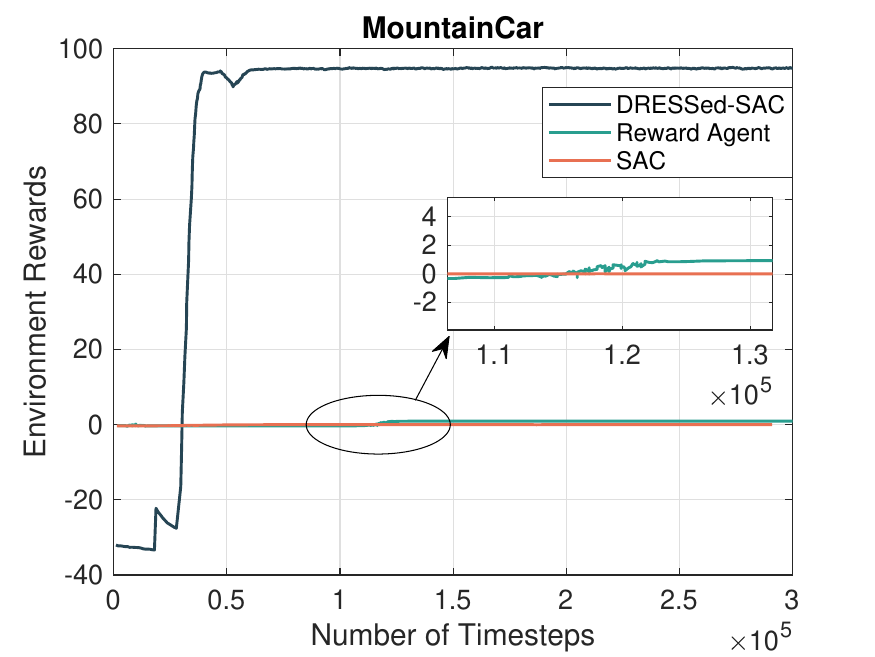}
}
\caption{Comparison of training performance in several DRL benchmark environments across DRESSed-SAC, reward agent-based method, and SAC.}
\label{fig:drlenv}
\end{figure*}

DRESS demonstrates strong generalizability in adapting various DRL architectures and diverse tasks. DRESS's ability to enhance different DRL algorithms has been discussed through Fig.~\ref{fig:comparedrl}. We further evaluate its task-generalizability across seven standard Gym control benchmarks: \textit{BipedalWalker}, \textit{AntStand}, \textit{AntFar}, \textit{HumanStand}, \textit{HumanKeep}, \textit{WalkerKeep}, and \textit{MountainCar}.

Table~\ref{tablec} compares DRESSed-SAC against three specialized reward shaping algorithms (Reward Agent, ROSA, ExploRS) and standard DRL implementations. The results show that DRESSed-SAC achieves superior performance in six out of seven environments, with only \textit{HumanStand} showing competitive results from the reward agent-based method. 
Figs.~\ref{fig:drlenvs} and \ref{fig:drlenv} provide detailed comparisons of training performance across these diverse DRL benchmark environments, highlighting DRESS's consistent ability to improve both learning speed and final performance. Specifically, Fig.~\ref{fig:drlenvs} examines the \textit{BipedalWalker} environment, where a reward of $300$ represents the threshold for successfully solving the task. In this challenging environment, DRESSed-SAC reaches this threshold within approximately $250,000$ timesteps, while standard SAC plateaus around $200$ even after $300,000$ timesteps. DRESSed-SAC achieves rewards $200\%$ higher than its basic version and shows a $33\%$ improvement over recent reward agent-based shaping methods.

\section{Conclusion}\label{Cons}
We proposed a novel reward shaping scheme, DRESS, to address challenges in robust network optimization for modern wireless communication systems. 
DRESS leverages the reasoning capabilities of diffusion models to infer reward signals from network states and actions, enabling more effective and robust DRL training in dynamic and complex wireless networks. 
Experimental results demonstrated that DRESS significantly improves the learning efficiency and performance of various DRL algorithms across diverse optimization problems, from resource allocation to motion control tasks. 
In our wireless benchmark environment, DRESSed-DRL achieved about $1.5{\rm x}$ times faster convergence than its original version in sparse-reward scenarios. 

The proposed DRESS-based DRL framework represents a practical solution to the evolving challenges of 6G wireless networks, providing a robust and efficient approach for optimizing network performance under extreme conditions. 
Future work could extend DRESS to additional wireless domains, such as channel knowledge graph generation, integrated sensing and communication, and network security applications, while further refining its integration capabilities to accommodate the unique characteristics of specialized communication networks.

\bibliographystyle{IEEEtran}
\bibliography{Ref}

\end{document}